\documentclass{aa}  

\usepackage[para]{threeparttable}
\usepackage{natbib}
\usepackage[colorlinks=true, allcolors=blue]{hyperref}
\usepackage{graphicx}
\usepackage{txfonts}
\usepackage{subcaption}         
\usepackage{lscape}             
\usepackage{placeins}           
\nolinenumbers
\def\Tstar{2900}
\def\Tdisk{600}
\def\Av{1.5}
\def\logg{3.0}

\begin{document}

   \title{PDRs4All}
   \subtitle{XVIII. JWST-NIRCam photometric properties of protoplanetary disks in the Orion Nebula Cluster}

   \author{P. Amiot\inst{1}
          \and O. Berné\inst{1}
          \and I. Schroetter\inst{1}
          \and M. Robberto\inst{2, 3} 
          \and T. J. Haworth\inst{4} 
          \and C. Boersma\inst{5}
          \and E. Dartois \inst{6}
          \and A. Fuente\inst{7}
          \and J.R. Goicoechea\inst{8}
          \and E. Habart\inst{9}
          \and M. J. McCaughrean\inst{10}
          \and T. Onaka\inst{11}
          \and E. Peeters\inst{12,13}
          }

   \institute{Institut de Recherche en Astrophysique et Planétologie, Université de Toulouse, Centre National de la Recherche Scientifique (CNRS), Centre National d’Etudes Spatiales, 31028 Toulouse, France \\ \email{paul.amiot@gmail.com}  
   \and
   Space Telescope Science Institute, 3700 San Martin Dr., Baltimore, MD 21218, USA 
   \and
   Johns Hopkins University, Bloomberg Center for Physics and Astronomy, 3400 N. Charles Street, Baltimore, MD
   21218, USA 
   \and
   Astronomy Unit, School of Physics and Astronomy, Queen Mary University of London, London E1 4NS, UK  
   \and
   NASA Ames Research Center, MS 245-6, Moffett Field, CA 94035-
   1000, USA 
   \and
    Institut des Sciences Moléculaires d’Orsay, Université Paris-Saclay, CNRS, Bâtiment 520, F-91405 Orsay Cedex, France
   \and
   Centro de Astrobiología (CAB), INTA-CSIC, Carretera de Ajalvir Km. 4, Torrejón de Ardoz, 28850 Madrid, Spain
   \and 
   Instituto de Física Fundamental (CSIC), Calle Serrano 121, 28006, Madrid, Spain
    \and
    Institut d’Astrophysique Spatiale, Université Paris-Saclay, CNRS, Bâtiment 121, F-91405 Orsay Cedex, France
    \and
    Max-Planck-Institute for Astronomy, K\"onigstuhl 17, 69117 Heidelberg, Germany
    \and
    Department of Astronomy, Graduate School of Science, The University of Tokyo, 7-3-1 Bunkyo-ku, Tokyo 113-0033, Japan 
   \and
   Department of Physics and Astronomy, University of Western Ontario, London, Ontario N6G 2V4, Canada
   \and 
   Institute for Earth and Space Exploration, University of Western Ontario, London, Ontario N6A 5B7, Canada
   }

   \date{Received 12 December 2025 / Accepted 16 February 2026}

  \abstract
   {
   The Orion Nebula Cluster (ONC) provides the closest example of ongoing star and planet formation in highly irradiated environments. 
   In particular, it is a key region for studying how ultraviolet (UV) radiation  from massive stars can drive mass loss in protoplanetary disks through photo-evaporation. Far-UV (FUV, energy $6<E<13.6$~eV) photons heat up the gas of the disk, forming a wind of neutral gas that is a photodissociation region (PDR). 
   We used high-angular-resolution NIRCam images from the \textit{PDRs4All} program and combined them with those of the guaranteed time observation (GTO) program 1256. 
   From these images, we extracted key information on ONC disks, such as the radii of the disks observed in silhouette against the bright background, the presence and positions of the dissociation fronts (DFs), the presence and positions of ionization fronts (IFs), intensities of Paschen $\alpha$ lines, and their near-infrared spectral energy distributions (SEDs).
   From this information we constructed a typology for ONC disks: {\it Type I} sources show an IF and DF nearly merged at the disk surface; {\it Type II} sources have their DFs at the disk surface and IFs at a distance of several tens of astronomical units from the disk; and {\it Type III} sources also have their DF at the disk surface, but show no IF. For all types of disks, we find that PAH emission traces the PDR. 
   We established that the SEDs of candidate Jupiter-mass binary objects (JuMBOs) observed as part of the {\it PDRs4All} program are  similar to the SEDs of {\it Type III} ONC disks, except for one of them, JuMBO24, which is of {\it Type I} or {\it II}. A detailed look at this SED shows that it is compatible with a young low-mass binary star with an unresolved ionized disk: a microproplyd binary. 
   We observe that the disk radius of ONC disks, $r_{\rm disk}$, increases with increasing projected distance to the ionizing source, $d_{\rm proj}$, following a power law, $r_{\rm disk}\propto d_{\rm proj}^{0.30}$, which is interpreted as evidence of the truncation of the disks by the photoevaporation (as reported in previous studies).
   The disk radii measured at infrared wavelengths appear larger than the disk radii measured at millimeter wavelengths, which is interpreted as evidence of the dust radial segregation within the disks.
   In agreement with theoretical models and observations of PDRs in the interstellar medium, the thermal pressure within the PDRs of ONC disk increases with the intensity of the FUV radiation field, $G_0$, but with a flatter slope.
}

\titlerunning{ONC disks with JWST}
   \maketitle

   \nolinenumbers

\section{Introduction}

\begin{figure*}
    \centering
    \includegraphics[width=0.9\linewidth]{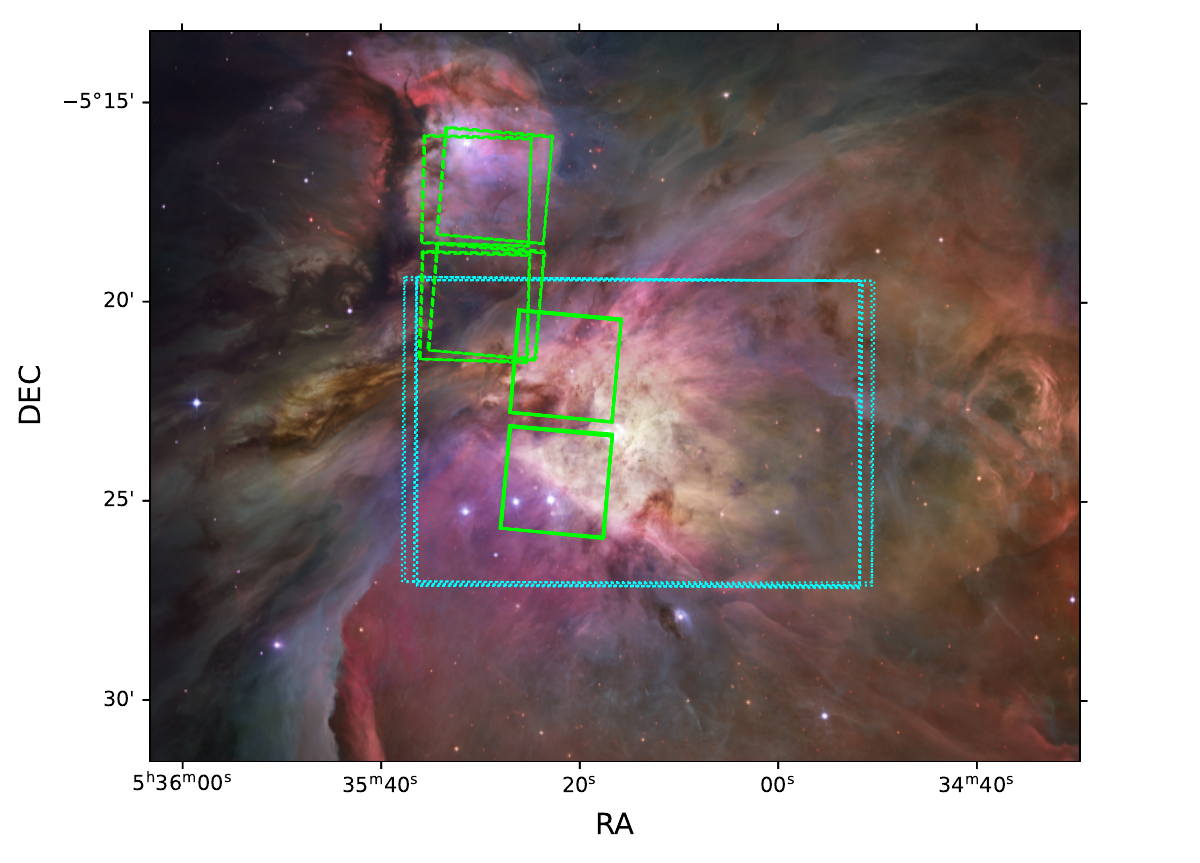}
    \caption{
    Composite color Hubble Space Telescope view of the Orion Nebula as provided by \citet{robberto_2013_HST}. The FOV of the NIRCam images from GTO program 1256 is shown as the dotted cyan box. Both FOVs from the \textit{PDRs4All} NIRCam images are shown as the continuous boxes, and the parallel FOVs are shown with dashed boxes for each of the filters listed in Table~\ref{tab:filters}.
    }
    \label{fig:ONC_FOV}
\end{figure*}

\begin{table*}[t]
    \caption{NIRCam filters used for the GTO 1256 and ERS-\textit{PDRs4All} observation programs of the ONC. }
\centering
  
    \begin{tabular}{cccrlcccc}
    \hline \hline 
Filters & $\lambda$ pivot & bandwidth & \multicolumn{2}{c}{Spatial resolution\tablefootmark{1}} & noise level\tablefootmark{3} &  \multicolumn{3}{c}{dataset} \\ \cline{7-9} 
        & (\textmu{}m)    &  (\textmu{}m)  & (arcsec) &  (au)\tablefootmark{2}  &  (W~m$^{-2}$\textmu{}m$^{-1}$sr$^{-1}$) &  GTO &   ERS  & ERS$_{\text{parallel}}$   \\ \hline 
F115W  & 1.154 &  0.225 & 0.040 & 15.6 & 3.1$\times10^{-6}$       &  $\surd$ &   &    \\   
F140M  & 1.404 &  0.142 & 0.048 & 18.7 & 1.6$\times10^{-6}$   &  $\surd$ & $\surd$ &    \\   
F162M  & 1.626 &  0.168 & 0.055 & 21.4 & 1.3$\times10^{-6}$   &  $\surd$ & $\surd$ &    \\   
F164N  & 1.644 &  0.020 & 0.056 & 21.8 & 8.4$\times10^{-6}$   &    & $\surd$ &    \\   
F182M  & 1.845 &  0.238 & 0.062 & 24.2 & 1.9$\times10^{-6}$   &  $\surd$ & $\surd$ & $\surd$  \\   
F187N  & 1.874 &  0.024 & 0.064 & 25.0 & 1.1$\times10^{-5}$   &  $\surd$ & $\surd$ & $\surd$  \\   
F210M  & 2.093 &  0.205 & 0.071 & 27.7 & 8.1$\times10^{-7}$   &    & $\surd$ & $\surd$  \\   
F212N  & 2.120 &  0.027 & 0.072 & 28.1 & 4.4$\times10^{-6}$   &  $\surd$ & $\surd$ & $\surd$  \\   
F277W  & 2.786 &  0.672 & 0.092 & 35.9 & 4.0$\times10^{-7}$   &  $\surd$ & $\surd$ &    \\   
F300M  & 2.996 &  0.318 & 0.100 & 39.0 & 3.7$\times10^{-7}$   &  $\surd$ & $\surd$ & $\surd$  \\   
F323N  & 3.237 &  0.038 & 0.108 & 42.1 & 2.2$\times10^{-6}$   &    & $\surd$ &    \\   
F335M  & 3.365 &  0.347 & 0.111 & 43.3 & 2.3$\times10^{-6}$   &  $\surd$ & $\surd$ & $\surd$  \\   
F360M  & 3.621 &  0.372 & 0.120 & 46.8 & 3.3$\times10^{-7}$   &  $\surd$ &   &    \\   
F405N  & 4.055 &  0.046 & 0.136 & 53.0 & 1.8$\times10^{-6}$   &    & $\surd$ & $\surd$  \\   
F410M  & 4.092 &  0.436 & 0.137 & 53.4 &                  &    &   & $\surd$  \\   
F444W  & 4.421 &  1.024 & 0.145 & 56.5 & 2.8$\times10^{-7}$                 &  $\surd$ &   &    \\   
F470N  & 4.707 &  0.051 & 0.160 & 62.4 & 6.4$\times10^{-7}$   &  $\surd$ & $\surd$ &    \\   
F480M  & 4.834 &  0.303 & 0.164 & 64.0 & 3.9$\times10^{-7}$   &    & $\surd$ &    \\  
\hline
    \end{tabular}

    \label{tab:filters}
    \tablefoot{
ERS$_{\rm parallel}$ describes the filters used in the NIRCam parallel observation.\\
\tablefoottext{1}{Full width at half maximum of NIRCam PSFs.}
\tablefoottext{2}{Considering a distance to the ONC of 390 pc. }
\tablefoottext{3}{Standard deviation estimated in uniform regions of the images using a circular aperture of radius 3'', from the ERS-\textit{PDRs4All} images when available, and otherwise from GTO data.}
}
\end{table*}

Most of the stars and planetary systems form in clusters alongside massive stars \citep{lada_embedded_2003}. This includes the Solar System \citep{adam_2010}. 
Thus, to properly study the early evolution of stars and planetary systems one should focus on typical star-forming regions that are irradiated by ultraviolet (UV) photons emitted by massive stars. 
One of the closest representatives of such a region is the Orion Nebula, located at a distance of 390 pc \citep{maiz_apellaniz_villafranca_2022}. It is illuminated by a compact group of massive stars, the Trapezium, whose main UV source is $\theta^1$ Ori C \citep{simondiaz_2006_trapezium}. 
The first images of this region with the Hubble Space Telescope (HST) enabled the discovery of a wide variety of objects such as brown dwarfs, young stellar objects (YSOs), protoplanetary disks, and Herbig-Haro objects \citep{odell_discovery_1993, odell_1994_proplyds, McCaughrean1996, luhman_initial_2000, bally_2000}. 
One of the most striking discoveries was the existence of {\sl proplyds} \citep{odell_discovery_1993}, which are young stars surrounded by a protoplanetary disk embedded in a cocoon of neutral gas, surrounded by a bright teardrop-shaped ionization front (IF) pointed toward the UV source.
This discovery stimulated theoretical work that proposed that the morphology of proplyds is due to the external photoevaporation process, triggered by the UV radiation from the massive stars in the Orion Nebular Cluster \citep[ONC;][]{johnstone_1998, storzer_photodissociation_1999, adams_photoevaporation_2004}. The far-UV (FUV) photons (photons with energies of $6<E<13.6$~eV) reach the disk surface and heat and dissociate molecules, producing a photodissociation region (PDR). The thermal temperature of the gas in the PDR exceeds the escape velocity, 
causing the gas to be removed from the disk in the form of a thermal wind that is then ionized by the extreme-UV (EUV) photons (photons with $E>13.6$~eV).
This process removes material from the disks, impacting their total mass and the potential of them to form planets. 
Observations of the ONC at submillimeter wavelengths indeed show that the disks exposed to harsh UV fields are dust-mass depleted. Less massive disks are found closer to the Trapezium \citep{mann_submillimeter_2010, mann_alma_2014}, and the disks are more compact than those in low-mass star-forming regions, such as Lupus and Taurus \citep{eisner_protoplanetary_2018}.

The James Webb Space Telescope \citep[JWST,][]{Gardner_2023_JWST} is now offering unprecedented views of star-forming regions and proplyds at infrared (IR) wavelengths. 
In particular, spectroscopic observations of two ONC disks (203-504 and 203-506) demonstrated the crucial role of FUV photons in disk chemistry  \citep{berne_formation_2023, schroetter_2025_506, Zannese2025_506}. These observations also enabled the characterization of the gas in the disk 203-506 and provided an estimate of the mass-loss rate \citep{berne_far-ultravioletdriven_2024}.
However, to date, no study has yet taken advantage of the images captured with NIRCam, which provides high-spatial-resolution ($\rm \approx 0.03''/pix$) near-IR images \citep{Rieke_2023_NIRCam}.
At this time, two programs have targeted the ONC with NIRCam: the \textit{PDRs4All} Early Release Science (ERS) program 1288 \citep{berne2022_PDRs4All}, which provides 14 NIRCam images with narrow- and broadband filters \citep{habart_2024_NIRCam} of the Orion Bar and the north of the Dark Bay, and guaranteed time observation (GTO) program 1256 \citep{mccaughrean_jwst_2023}, which targeted the inner part of the ONC with 12 narrow and broadband filters.
A detailed analysis of the \textit{PDRs4All} NIRCam images can be found in \citet{habart_2024_NIRCam}, with a focus on the interstellar medium (ISM) and the structure of the nebula itself. The analysis of the disks and YSOs relied only on the F187N filter.
A description of the GTO 1256 NIRCam images is provided in \citet{mccaughrean_jwst_2023}. 
Those authors briefly describe the observational results obtained for the Nebula, brown dwarfs, and some protoplanetary disks. 
In these studies, no statistical analysis of the disks or the effect of photoevaporation was performed. 

Analyzing the images from GTO 1256, \citet{pearson_jupiter_2023} reported the presence of 540 faint sources, which they interpreted as free-floating planetary mass objects. 
This sample contains 40 binary and two triple systems, referred to as JuMBOs (Jupiter-mass binary objects). The existence of such a number of pairs has motivated theoretical work aimed at understanding their formation mechanisms \citep{wange_2024_jumbo_formation, diamond_formation_2024}. However, the nature of these sources remains controversial.
\citet{luhman_candidates_2024} performed a counter-analysis of the NIRCam images that challenges the planetary nature of JuMBOs and suggests that some of them are in fact background stars. JWST-NIRSpec spectroscopic observations support this idea for seven out of 40 JuMBOs \citep{luhman_2025_spectra}. Further NIRSpec observations will provide a better assessment of the nature of these sources (McCaughrean et al., in prep).
Analyzing archival observations with the Karl G. Jansky Very Large Array (VLA) of the ONC, \citet{rodriguez_radio_2024} found a radio counterpart for one of the JuMBO, JuMBO24\footnote{source 274 in \citet{luhman_initial_2000}, source 728 in \citet{Slesnick_2004}}. 
They interpreted this emission as being due to the possible presence of a radiation belt or aurora. The nature of this radio emission is still unclear, however.
Currently, the evidence of JuMBOs remains limited to the GTO 1256 images.

In this paper, we make use of the \textit{PDRs4All} data and its suite of NIRCam filters to analyze, in complement to the GTO 1256 data, the properties of disks in ONC and JuMBOs. 
Our objectives are to assess the photometric properties of irradiated protoplanetary disks within the ONC, relate the properties to the physical parameters of those disks and the effect of photoevaporation, and use these insights to assess the characteristics of JuMBOs.

In Section~\ref{sect:observations}, we describe the data and the sample of disks and JuMBOs. 
In Section~\ref{sect:images_ONCdisks}, we describe the NIRCam images of the disks and JuMBOs, and build images of the disks in specific emission lines; namely, [\ion{Fe}{ii}], Paschen $\alpha$, H$_2$ 1-0 S(1), and PAHs (3.3 \textmu{}m). 
In Section~\ref{sect:proplyd_or_not}, we derive a typology based on the emission maps and relate it to the spectral properties of the disks and JuMBOs.
In Section~\ref{sect:NIRproperties}, we investigate the NIR properties of the disks, such as the disk and IF radius and the Paschen $\alpha$ intensity, and compare our results with previous studies.
Lastly, in Section~\ref{sect:ONCdisk_PDR}, we relate the properties of the ONC disks to the FUV field, and to the properties of PDRs.

\section{Observations, data, and catalog}\label{sect:observations}

\subsection{JWST NIRCam data}\label{sect:data}

We used the JWST NIRCam images of the ONC from the \textit{PDRs4All} ERS program \citep[][PID 1288]{berne2022_PDRs4All} and GTO program 1256 \citep{mccaughrean_jwst_2023}. 
The fields of view for these two programs are presented in Fig.~\ref{fig:ONC_FOV} and the selected NIRCam filters in Table~\ref{tab:filters}. 
The images used for this paper were reduced using the procedure described in 
\citet{habart_2024_NIRCam} and \citet{luhman_candidates_2024} for the \textit{PDRs4All} and 
GTO 1256, respectively. The \textit{PDRs4All} data are available through the specific \textit{PDRs4All} MAST archive\footnote{\url{https://mast.stsci.edu/portal/Mashup/Clients/Mast/Portal.html}} and through the REGARDS database\footnote{\url{https://regards.osups.universite-paris-saclay.fr/user/jwst}}.
The GTO 1256 are available here\footnote{\url{https://zenodo.org/records/13824020}}. 
The \textit{PDRs4All} dataset contains two $\approx$2.5'$\times$2.5' fields of view (FOVs; Fig.~\ref{fig:ONC_FOV}), with 14 filters (Table~\ref{tab:filters}) covering the north of the Dark Bay, to the northeast of the Trapezium Cluster (module A) and the Orion Bar, to the southeast of the Trapezium Cluster (module B).  
This dataset also contains parallel observations in eight filters (Table~\ref{tab:filters}) covering the M43 (NGC 1982) region, situated to the northeast of the Orion Bar. 
The GTO 1256 program provides a 11'$\times$7.5' mosaic (Fig.~\ref{fig:ONC_FOV}) in 12 filters (Table~\ref{tab:filters}) centered on the Trapezium cluster. {The pixel size of the NIRCam images is 0.031''/pix in the short wavelength channel (0.7-2.3~\textmu{}m),  and 0.064’’/pix in the long wavelength channel (2.3 - 4.8~\textmu{}m)}.

\subsection{Disk and JuMBO catalog}\label{sect:sample}

To construct our sample of ONC disks, we refer to the catalog of \citet{ricci_hubble_2008}. It contains 219 sources detected with the Hubble Space Telescope (HST) that show evidence of circumstellar matter. The \citet{ricci_hubble_2008} catalog relies on images obtained with the Wide Field Channel of the Advanced Camera for Surveys of the ONC as part of the Treasury Program on the Orion Nebula \citep{robberto_2013_HST}.
In the JWST NIRCam data (combining all filters from both GTO and ERS programs), we identify emission from 188 sources of the \citet{ricci_hubble_2008} sample. These sources are listed in Table \ref{tab:coordinates}, named as in \citet{ricci_hubble_2008}. 
We also include additional sources not reported by \citeauthor{ricci_hubble_2008}   
This includes the three new proplyd candidates reported by \citet{habart_2024_NIRCam} in the \textit{PDRs4All} images (171-212, 180-218, and 234-104, see Fig.~\ref{fig:HC182}). In addition, we identified in the NIRCam images a source (HC182 in the catalog of \citealt{hillenbrand_constraints_2000}) that exhibits a disk seen in silhouette against the background and a bright cocoon in the F187N filter (Fig.~\ref{fig:HC182}).
This source is included in our catalog. 
Overall, our sample of protoplanetary disks contains 192 sources, whose coordinates are given in Table~\ref{tab:coordinates}.

A catalog and analysis of JuMBOs using the GTO 1256 NIRCam images has been presented in  \citet{pearson_jupiter_2023}. {The spectral energy distribution (SED) of the JuMBOs in the GTO 1256 images are provided in the latter work.} 
To complement this analysis, we rely on the \textit{PDRs4All} NIRCam images.
{Due to their smaller FOV (see Fig.~\ref{fig:ONC_FOV}),  the \textit{PDRs4All} images only cover 11 JuMBOs, whose names and coordinates are given Table~\ref{tab:coordinates_JuMBOs}}.
We note that JuMBO25, pertaining to this list, is one of the two triple system reported by \citet{pearson_jupiter_2023}.

\begin{figure*}[!ht]
    \centering
    \includegraphics[width = 0.85\linewidth]{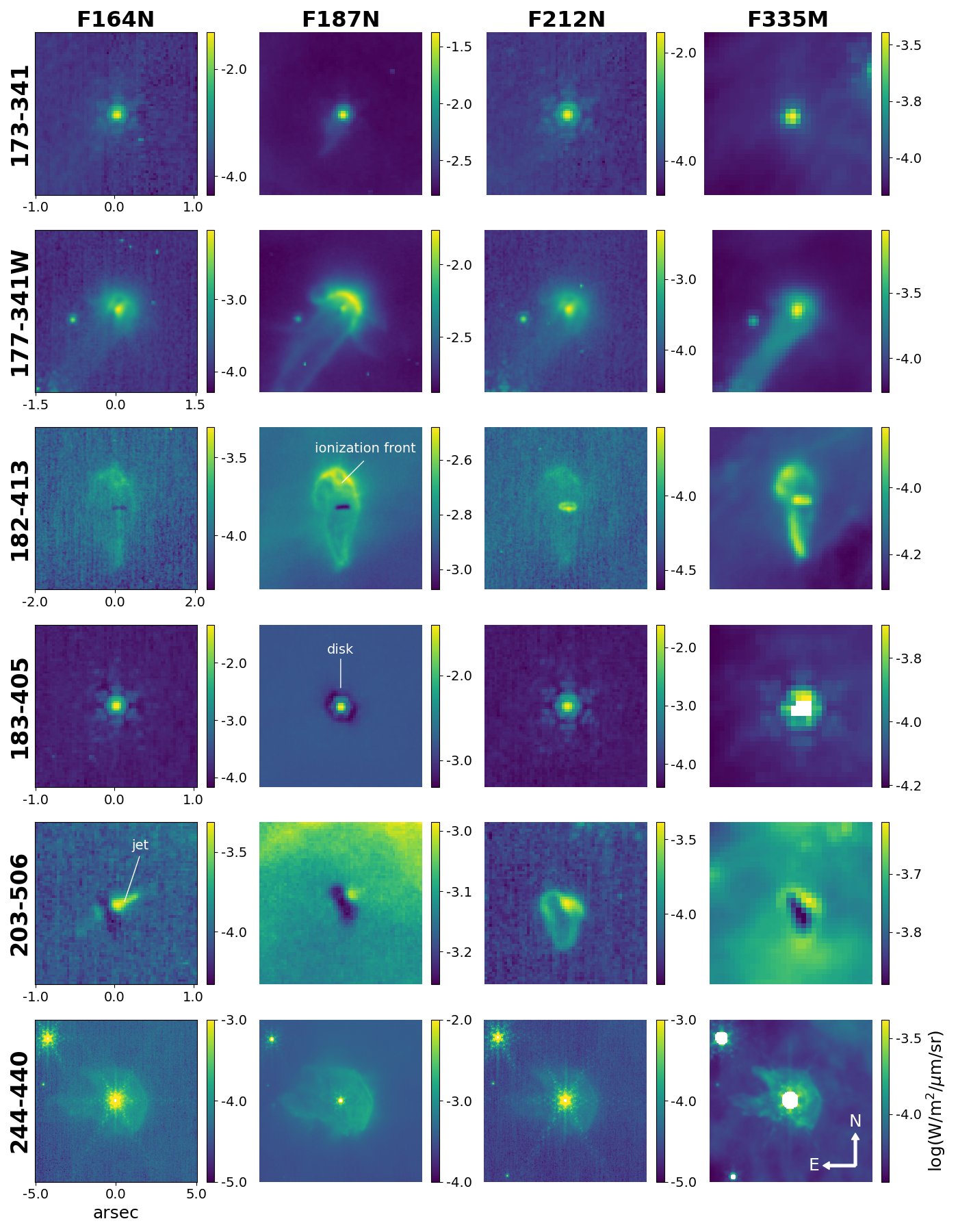}
    \caption{NIRCam images of the disks seen in silhouette in the narrowband filters from the \textit{PDRs4All} program dataset. The color scale is in $\log (\rm W\:m^{-2}$ \textmu m$^{-1}$ sr$^{-1})$.}
    \label{fig:kaleidoscope_proplyds}
\end{figure*}

\section{Images of ONC disks and JuMBOs}\label{sect:images_ONCdisks}

\subsection{ONC protoplanetary disks}\label{sect:PPDs}

Fig.~\ref{fig:kaleidoscope_proplyds} shows the NIRCam images of six protoplanetary disks, chosen to illustrate the diversity of sources present in the ONC. 
An important difference between the JWST observation of these disks vs. those with HST is that the central stars are bright and sometimes dominate and/or saturate the image, especially at the longest wavelengths.
This is a natural result of the lower extinction and brighter flux of these young low-mass objects in the IR with respect to the visible.

\begin{figure}
\centering    
     \includegraphics[width=\linewidth]{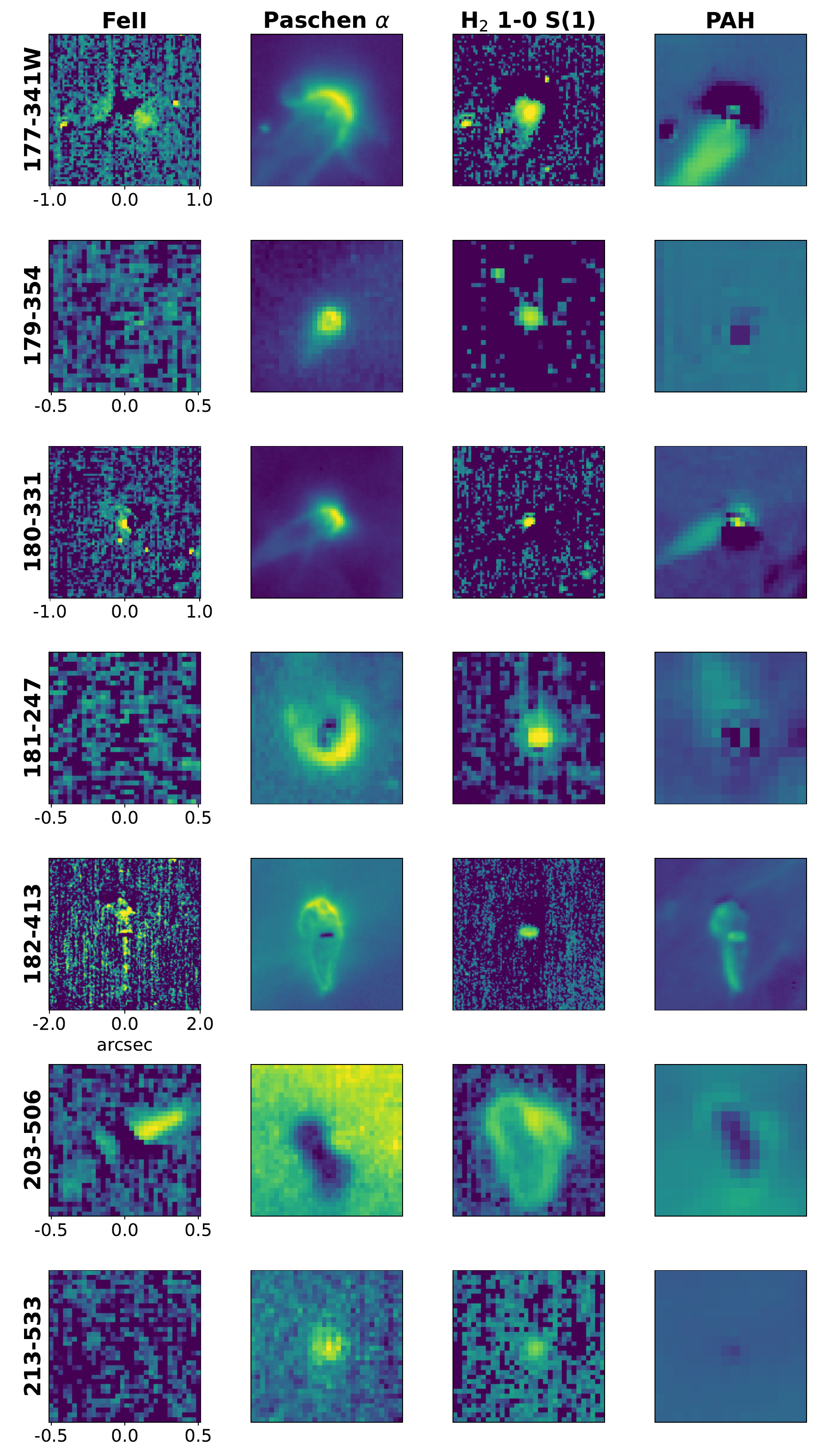}
     \caption{[\ion{Fe}{ii}], Paschen $\alpha$, H$_2$ 1-0 S(1) and PAH emission maps of a subsample of ONC protoplanetary disks, obtained following Equation~\ref{eq:recipes_lines}. The images are north-aligned and in units of watt per square meter per steradian.} 
     \label{fig:map_lines0}

\end{figure}

In Fig.~\ref{fig:kaleidoscope_proplyds}  we present images in four filters -- F164N, F187N, F212N, and F335M -- to illustrate the general morphology of ONC disks as seen with the JWST. 
In the F164N filter, corresponding to the [\ion{Fe}{ii}] emission line, emission mostly arises from the central source and, in some cases, also from a jet.
This is clearly the case for 203-506 \citep[see][]{berne_far-ultravioletdriven_2024}, but also for 182-413, although much less clearly. 
In the F187N filter, the disks surrounding the young stars are seen in silhouette against the bright background (e.g., 177-341W, 182-413, 183-405, and 203-506). 
The F187N filter, corresponding to the Pa $\alpha$ line,  provides a striking view of the bright cocoons and IFs surrounding some of the disks in ONC, similar to the HST images in the H$\alpha$ line \citep[e.g.,][]{bally_2000, ricci_hubble_2008}.
In Fig.~\ref{fig:map_lines0}, this is mostly evident for 177-341W, 182-413, and 244-440, where the teardrop-shaped cocoon is particularly bright relative to the background. 
In the F212N filter, corresponding to the 1-0~(S1) H$_2$ line, only the star is seen in general. However, in some cases, the disk surface is seen in emission (203-506 and 182-413).
The emission in the F335M filter, encompassing both PAH and CO$_2$ lines, appears to arise from the disk surface and from the PDR. 
For 182-413, this was previously observed with the Very Large Telescope/VISIR instrument in a PAH filter \citep{vicente_2013_HST10} and with Keck/NIRC2 at 3.3~\textmu{}m \citep{shuping_2014_HST10}.

The [\ion{Fe}{ii}] line at 1.64 \textmu{}m traces hot ionized gas. In environments similar to those found in the ONC disks, this is a good tracer of shocks. 
The Paschen $\alpha$ recombination line at 1.87~\textmu{}m traces the recombination of hydrogen at the IF. The 2.12~\textmu{}m line of molecular hydrogen (H$_2$ 1-0 S(1)) traces the dissociation front (DF), which is the interface between atomic and molecular gas in far-UV irradiated environments \citep{berne_far-ultravioletdriven_2024, schroetter_2025_506, Goicoechea2025}. Finally, the emission of polycyclic aromatic hydrocarbon (PAH) at 3.3~\textmu{}m traces the C-H in-plane stretch emission \citep[e.g.,][]{Peeters2024_PDRs4All, schroetter_2025_506}.

In order to extract maps tracing different physical conditions from the NIRCam images, we relied on the empirical prescriptions of \citet{Chown2025_calibration}. These authors provide recipes to compute the intensity in specific lines ($I_{\rm line}$) and bands from a combination of NIRCam filters with equations of the form: 
\begin{equation}\label{eq:recipes_lines}
I_{\rm line} = (F_{\rm \lambda}^{\rm filtA} -\alpha\:F^{\rm \rm filtB}_{\lambda})\times BW_{\rm filtA}+\beta,
\end{equation}
where $BW_{\rm filter}$ and $F_{\lambda}^{\rm filter}$ are the bandwidth (cf. Table~\ref{tab:filters}) and the flux density in the corresponding filter, respectively, and $\alpha$ and $\rm \beta$ the coefficients retrieved by \citet{Chown2025_calibration}. Here “{\sl filtA}” corresponds to the filter capturing the emission of interest and "{\sl filtB}" is the filter used to subtract continuum emission. The filters
combination and the coefficients are given in Table~\ref{tab:recipes_line}.

\begin{table}[h]
\caption{Filter combinations and coefficients for Equation~\ref{eq:recipes_lines} for selected lines from \citet{Chown2025_calibration}. $\alpha$ is unitless and $\beta$ is in watts per square meter per steradian.}
\centering
\begin{tabular}{r l c c c c}
\hline\hline
\multicolumn{2}{c}{Line (\textmu{}m)} & Filter A & Filter B & $\alpha$ & $\beta$ \\ \hline
[\ion{Fe}{ii}]  & (1.64) & F164N & F162M & 1.10 & $8.3 \times 10^{-8}$ \\
Pa~$\alpha$     & (1.88) & F187N & F182M & 0.72 & $-1.3 \times 10^{-7}$ \\
H$_2$           & (2.12) & F212N & F210M & 0.78 & $1.5 \times 10^{-8}$ \\
PAH             & (3.3)  & F335M & F300M & 1.13 & $2.0 \times 10^{-6}$ \\
\hline
\end{tabular}
\label{tab:recipes_line}
\end{table}

Excluding those that are saturated in more than two filters, we are left with a sample of 22 sources. Their [\ion{Fe}{ii}], Paschen $\alpha$, H$_2$ 1-0 S(1), and PAH maps are presented in Appendix \ref{sect:All_maps}. Fig.~\ref{fig:map_lines0} shows the map for a subsample chosen to illustrate the main observational results.

\subsection{JuMBOs as seen in the \textit{PDRs4All} data}

Fig.~\ref{fig:kaleidoscope_JuMBOs} shows the NIRCam images of six JuMBOs in the F140M, F187N, F210M, and F277W filters from the \textit{PDRs4All} dataset. 
We present mainly broadband filters in this figure, since they facilitate the detection of the faintest sources (such as JuMBOs), but we also include the 
F187N as a comparison narrow-band filter. 
In Fig.~\ref{fig:kaleidoscope_JuMBOs}, we see that the brightest source from each JuMBO is detected in most filters. In the F187N filter, however, the sources are hardly visible (see JuMBO25 and JuMBO30) and are dominated by bright nebular emission.
The companions are mostly visible in the F210M and F277W filters. This is also the case for JuMBO25, which is a projected triple system. 
The images of JuMBO24 highlight how the increased angular resolution at shorter wavelengths allows one to resolve the two sources, separated by $\gtrsim 20$au.

\subsection{Herbig Haro object}

\begin{figure}[h]
    \centering
    \includegraphics[width=\linewidth]{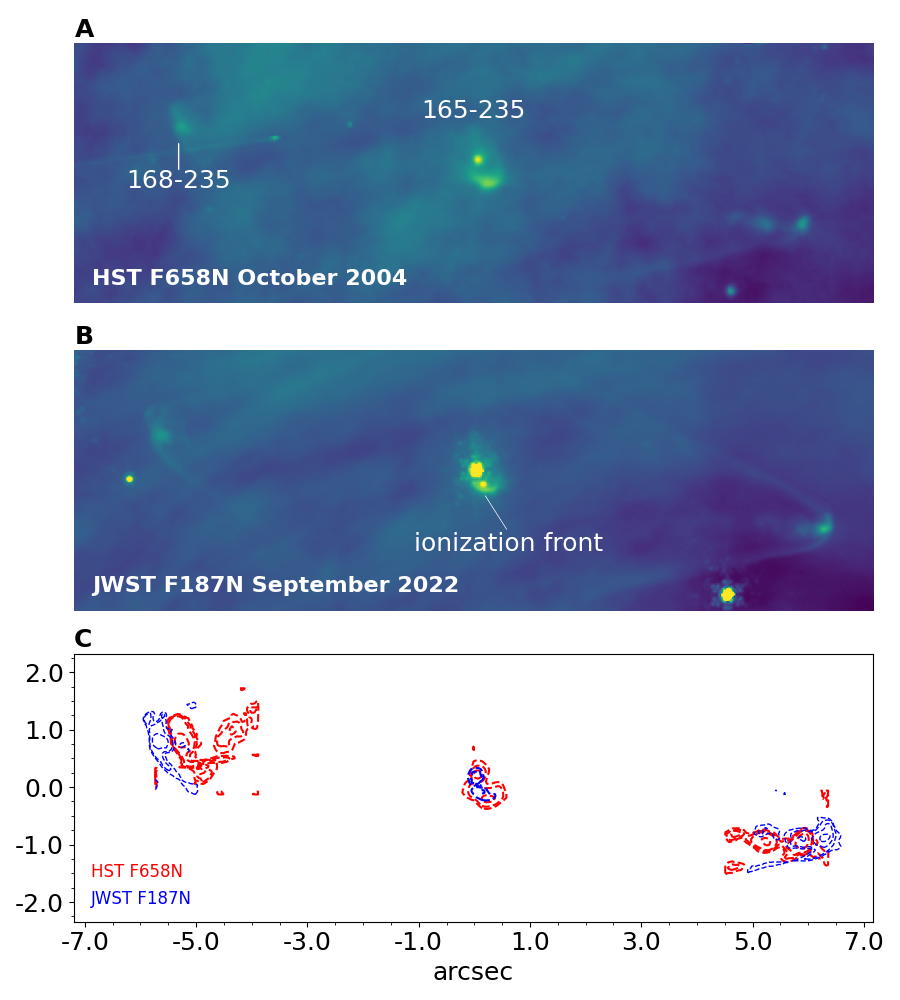}
    \caption{(\textbf{A}): HST-image of 165-235 in the H$\alpha$ F658N filter at 0.658~\textmu{}m. (\textbf{B}) JWST-image in the Pa $\alpha$ F187N filter at 1.87~\textmu{}m. (\textbf{C}): In red, contours of 165-235 from the HST image. In blue, contours from the JWST image.}
    \label{fig:HH513}
\end{figure}

168-235 was reported to be an irradiated disk by \citet{ricci_hubble_2008}.
However, this source was previously reported as a Herbig-Haro (HH513) resulting from a jet associated with 165-235 \citep{bally_2000}.
Fig.~\ref{fig:HH513} shows the HST F658N image taken October 2004, the F187N JWST NIRCam images taken September 2022, and 
the contours from both images combined.
Over nearly 18 years, the source has moved $\approx$0.5'' in the images, implying a velocity in the plane of the sky of $\approx25\:{\rm mas\:yr^{-1}}$. At a distance of 390 pc, this gives a projected velocity of $\approx 50\: {\rm km\:s^{-1}}$.
This high velocity together with the symmetry of a similar source at the opposite side of 165-235 confirms that 168-235 is a Herbig-Haro object rather than a protoplanetary disk. 
A bright bow shock-like structure associated with the central binary source is visible in the F187N/Pa $\alpha$ filter, but not visible in F212N/H$_2$, suggesting that this structure is an IF. 

\section{Typology of ONC disks}\label{sect:proplyd_or_not}

\subsection{Observational categories of ONC disks}\label{sect:typology}

\begin{figure*}
    \centering
    \includegraphics[width=\linewidth]{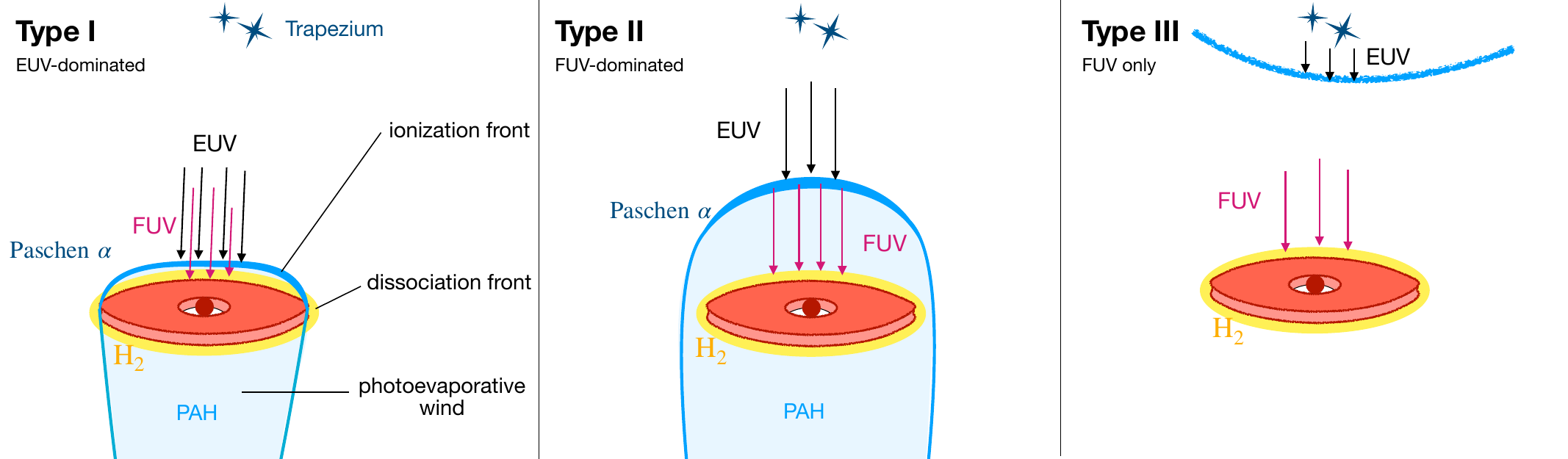}
    \caption{Schematic overview of the observational categories for ONC disks. Type I, the IF (Pa $\alpha$ emitting layer, in blue) is  close to the disk's surface and almost merged with the DF (H$_2$ emitting layer, in yellow).
     Type II, the IF is placed further away from the disk's surface and DF. 
     In these two first categories, the disk is in the \ion{H}{ii} region and embedded in a photoevaporative wind (PAH emitting area) surrounded by a bright IF.
    Type III, the disk does not have any associated IF, but still has a DF. This corresponds to the case where the disk is FUV irradiated only, beyond the Strömgren's sphere produced by the Trapezium.   
    }
    \label{fig:scheme_category}
\end{figure*}

Inspecting Fig.~\ref{fig:map_lines0} (and figures in Appendix~\ref{sect:All_maps}), we observe that the [\ion{Fe}{ii}] emission traces the jet solely (203-506, 177-341W, and 182-413) as the central star is never detected in the [\ion{Fe}{ii}] map.
In the Pa $\alpha$ map, the IF is particularly bright. We likely observe IF emission in 13 sources.
The Pa $\alpha$ maps also provide better contrast between the silhouette disks and the background (or the IF when present).
For five sources (177-341W, 180-331, 181-247, 182-413, and 183-419), the IF is distinct from the central star and disk surface. In these cases, we see that the central star does not significantly contribute to the Pa $\alpha$ emission when compared to the IF (see case for 181-247). 
For 175-355, 209-151, and 255-512 we do not see any silhouette disk, but the extended emission from the IF can be (roughly) distinguished from the central star. 
For the five remaining sources with extended emission, the tip of the IF is fully merged with the central star. 
We also observe bright arcs in two sources (180-218 and 234-104). However, it is not clear whether this arc is associated with the source, so we remain cautious regarding its nature. 
When no IF is observed, the central star remains bright in Pa $\alpha$ (e.g., 201-534, 239-334).

In the H$_2$ map, all central stars are bright. 
For 203-506 and 182-413 the emission also delineates the disk, indicating that the DF is located close to the disk's surface, as suggested by models of proplyds \citep{champion_2017_hershell}.
For 191-232, 218-529 and 239-334 the H$_2$ emission appears slightly extended.
Finally, the PAH emission maps generally trace the photoevaporative wind (see e.g., 177-341W, 180-331, 181-247, 183-419, but also 191-232), as was also seen in earlier studies \citep{vicente_2013_HST10, shuping_2014_HST10}.
For 182-413, 203-506, and 218-529, PAH emission traces the disk's surface, similarly to the H$_2$ emission. The combined presence of PAHs and H$_2$ emission in the near IR is typical of PDRs as seen with JWST \citep{Peeters2024_PDRs4All}. 

Based on these observational results, we classify the disk from our sample of 22 sources into three categories illustrated in Fig.~\ref{fig:scheme_category}. 
{\bf \textit{Type I}} (left panel) corresponds to  sources where the IF (Pa $\alpha$ emitting layer) and DF (H$_2$ emitting layer) are separated by  $\lesssim$ 25 au (the NIRCam's separation limit around 2~\textmu{}m), so that they cannot be disentangled in NIRCam images. 
{\bf \textit{Type II}} (middle panel) corresponds to the sources with an IF placed further away from the disk's surface, so that the DF and the IF are physically separated by several tens of au ($\gtrsim$ 25 au). 
Here, the FUV radiation dominates at the disk surface and the lower density wind is irradiated by EUV photons further out, which produces an IF that is detached from the disk.
{\bf  \textit{Type III}} (right panel) corresponds to the sources that do not show signs of an IF, but that still present a DF seen in H$_{2}$. This corresponds to the case where the disk is FUV irradiated only.  
We identify the category for each of the 22 sources in Table~\ref{tab:typology_disks}.
Two sources, 180-218 and 234-104, are marked as “ambiguous,” as we do see an extended Pa $\alpha$ emission similar to an IF, but we cannot assert if this emission is associated with the source. 

\begin{table}
    \caption{List of categorized ONC disks and their associated type (see text and Fig.~\ref{fig:scheme_category}).}
\centering
    \begin{tabular}{cccc}
    \hline
    source & ra & dec & type  \\ \hline \hline
    177-341W & 83.8236919 & -5.3947273 & Type II \\
    180-331 & 83.8252228 & -5.391897 & Type I \\
    181-247 & 83.8253908 & -5.379755 & Type II \\
    182-413 & 83.8259312 & -5.4037385 & Type II \\
    183-419 & 83.826318 & -5.4052583 & Type II \\
    179-354 & 83.8248805 & -5.3981947 & Type I \\
    171-212 & 83.8213621 & -5.3699739 & Type I \\   
    190-251 & 83.8293351 & -5.3807361 & Type I \\
    209-151 & 83.837536 & -5.3645253 & Type II \\
    250-439 & 83.8543114 & -5.4106944 & Type I \\
    255-512 & 83.8564085 & -5.4199612 & Type II \\
    218-529 & 83.8409785 & -5.4245821 & Type III \\
    232-453 & 83.8468152 & -5.4146892 & Type III \\
    239-334 & 83.8494959 & -5.392798 & Type III \\
    234-104 & 83.8473365 & -5.3510211 & ambiguous \\
    175-355 & 83.8231177 & -5.3986324 & Type II \\
    180-218 & 83.8251087 & -5.3717045 & ambiguous \\
    191-232 & 83.8297216 & -5.375387 & Type II \\
    201-534 & 83.8339867 & -5.4260637 & Type III \\
    203-506 & 83.8346666 & -5.4182083 & Type III \\
    213-533 & 83.8387402 & -5.4258708 & Type III \\
    216-541 & 83.840053 & -5.4279793 & Type III \\

    \hline
    \end{tabular}
    \label{tab:typology_disks}
\end{table}

\begin{figure}
    \centering
    \includegraphics[width=\linewidth]{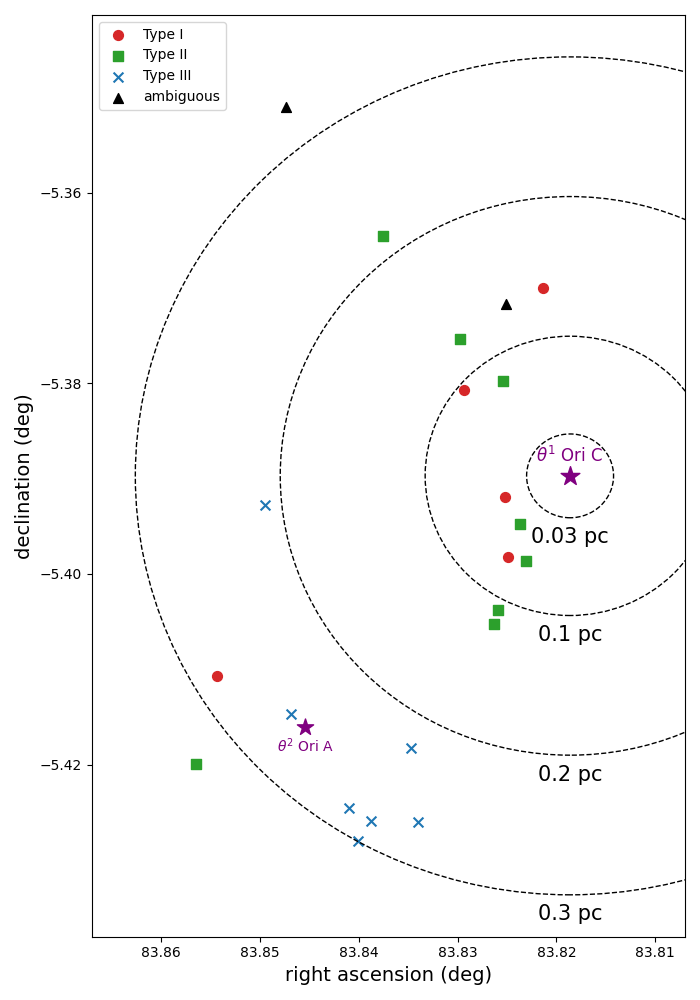}
    \caption{Spatial distribution of the 22 protoplanetary disks categorized in Section~\ref{sect:typology}. 
    The red dots are sources of {\it type I}. The green squares correspond to {\it type II} sources.
    The blue crosses are the {\it type III} sources.
    The dark triangles are ambiguous. 
    The position of $\theta^1$ Ori C and $\theta^1$ Ori A are marked with the purple stars. 
    244-440 and 203-504, which are illuminated by $\theta^1$ Ori A, are not in this subsample.}
    \label{fig:distribution_proplyds}
\end{figure}

Fig.~\ref{fig:distribution_proplyds} shows the projected spatial distribution of the sources in the ONC color-coded according the three groups. Some trends can be identified, in particular, beyond 0.2 pc from $\theta^1$ Ori C 70\% of the sources are {\it type III}. Interestingly, these sources appear to be concentrated around $\theta^2$ Ori A, which is the second dominant source of UV photons.
Instead, $90~\%$ of the sources within a 0.2 pc distance are of {\it type I} or {\it II}. This suggests that these three categories may be related to the intensity of EUV and FUV photons.  
In their models, \citet{johnstone_1998} describe two cases of photoevaporative flows. The {\bf “FUV-dominated flows”}, which occur for disks situated at a distance d>0.03~pc from $\theta^1$Ori~C. In this region, the IF is spatially separated from the disk at several 10s of au. The {\bf “EUV-dominated flows”}, which occur for sources within 0.03 pc from $\theta^1$ Ori C. In this regime the IF is merged with the disk surface.
It is tempting to associate our {\it Type II} sources presenting an IF separated from the disk's surface to the “FUV-dominated flows,” and the {\it Type I} sources to the “EUV-dominated flows,” as labeled in Fig.~\ref{fig:scheme_category}.
However, the dependence of our  {\it type I} - EUV-dominated flows and {\it type II} - FUV-dominated flows on the distance from $\theta^1$ Ori C is not clear and, in particular, we do not find any evidence of a clear separation between the two categories at a distance of 0.03 pc from $\theta^1$ Ori C, as suggested by \citet{johnstone_1998}. Some  {\it type I} - EUV-dominated flows are indeed seen at distances larger than  0.1 pc from $\theta^1$ Ori C. 
One has to remember that projected distances are just lower limits for the actual physical distances considered by \citet{johnstone_1998}. 
Moreover, \citet{storzer_photodissociation_1999, haworth2023_FRIEDv2} found that FUV should generally dominate in terms of setting the mass loss rate. In this case, close IF-DF spatial separation of {\it type I} sources could be due to smaller mass-loss rates, possibly due to smaller disks.
It is also possible for a disk to be seen in different stages if evolutionary effects (e.g., flyby to the inner regions of the nebula, dust depletion, etc.) play a role. Increasing the disk sample by adding the majority of disks that were excluded because of saturation would increase statistics and possibly shed light on this issue.

The categories we defined can also be discussed in the context of the term “Proplyd” \citep{odell_discovery_1993}, which is widely used. We follow the recent  definition given 
 in \citet{winter_external_2022}, {which defines a Proplyd as} “a circumstellar disk with an externally driven photoevaporative wind composed of a photodissociation region and an exterior ionization front with a cometary morphology.” 
 With this definition, {\it type I} and {\it II} disks fall in the category of Proplyds, and include some famous objects such as HST10 and 244-440. 
 Instead, dark silhouette disks such as 114-426 \citep{McCaughrean1996, miotello_evidence_2012,Ballering_2025_114-426}, the largest protoplanetary disk in the Orion Nebula, or
such as 203-506, widely studied with JWST recently \citep{berne_formation_2023, berne_far-ultravioletdriven_2024, schroetter_2025_506, Goicoechea2024_506, Zannese2025_506} are of {\it type III} and do not correspond to the most recent definition of Proplyd. However, it is important to note that these sources are known to still experience significant mass loss from FUV-driven photoevaporation
 \citep{miotello_evidence_2012, berne_far-ultravioletdriven_2024}.

\subsection{Spectral properties of ONC disks}

\subsubsection{Spectral energy distributions}\label{sect:SED}

For the photometric analysis, we rely exclusively on the \textit{PDRs4All} NIRCam images.
This reduces the main sample described in Section~\ref{sect:sample} to 85 protoplanetary disks (and 11 JuMBOs). 
To extract the photometry, we used the \texttt{AperturePhotometry} function from the \texttt{Photutils} python library\footnote{\url{https://photutils.readthedocs.io/en/stable/aperture.html}}. 
We used an aperture radius, $r_{\rm arsec}$, of 0.15'' centered on the source's coordinates. 
Images where this aperture encompasses more than two saturated pixels were excluded from the analysis. Otherwise (at most two saturated pixels), the values of saturated pixels were replaced by the maximum intensity observed in non-saturated pixels within the aperture.  
We estimated the background nebular emission in an annulus with $0.16$'' inner radius and 0.2'' width. 
Five sources have extended emission from the IF that extends beyond the 0.15'' aperture: 177-341W, 180-331, 181-247, 183-419, and 182-413.
For the first three, we considered an aperture radius of 0.3''. The background nebular emission was estimated in an annulus of $ 0.31$'' inner radius and  0.2'' width. 
Lastly, for 182-413 we used an elliptical aperture of 1.2'' and 0.6'' for the semimajor and semiminor axis, respectively. The background nebular emission was estimated in an annulus of 1.21'' inner radius and 0.2'' width. 
The source flux densities were given by the on-source flux density minus the background nebular emission estimated within the annulus. The photometric error on the source flux densities was given by three times the standard deviation of per-pixel flux density in the background annulus.
The sources flux densities were then corrected for encircled energy\footnote{\url{https://jwst-docs.stsci.edu/jwst-near-infrared-camera/nircam-performance/nircam-point-spread-functions\#gsc.tab=0}}.
We note that for the sources hosting a disk seen in silhouette, the flux densities obtained by this method in the Pa and Br filters are not reliable, as the nebular emission is brighter than the silhouette disk leading to an over-subtraction of the on-source flux. In this case, no fluxes are extracted.

From the determined photometry, we derived the SED of the disks and JuMBOs detected in the \textit{PDRs4All} NIRCam images. Among this sample, we discarded the SEDs for which there was saturation or non-detection in several filters. 
Finally, we obtained 23 SEDs for the protoplanetary disks and five for JuMBOs, which are presented in Fig.~\ref{fig:SED_proplyds}.

The major common characteristic of these SEDs is the decreasing slope with wavelength, as expected for Class II T Tauri stars.
Approximately 60\%
of the sources have SEDs with excess emission at 1.87~\textmu{}m and 4.05~\textmu{}m, i.e., the Pa and Br $\alpha$ lines. 
 In Fig.~\ref{fig:SED_proplyds}, we detect water ice absorption feature at 3.0~\textmu{}m (F300M filter) for 171-212, 180-218, 182-413, 183-419, and 203-506,  and with less confidence in 181-247 and 191-232. 
 PAHs emission at 3.3~\textmu{}m (F335M filter) is detected for 182-413, 183-419, 203-506 and likely in 191-232. 
 We likely detect $\rm H_2$ emission for five sources (180-218, 182-413, 183-419, 213-533, and 234-104), at least one of the following wavelengths: 2.12~\textmu{}m, 3.23~\textmu{}m, and 4.70~\textmu{}m  (filters F212N, F323N, and F470N). 
Lastly, we do not detect the forbidden iron lines [\ion{Fe}{ii}] at 1.64~\textmu{}m in our sample, as anticipated in Section \ref{sect:images_ONCdisks}.

\subsubsection{SEDs and ONC disk typologies}

\begin{figure*}[htb]
\centering
    \includegraphics[width=0.8\hsize]{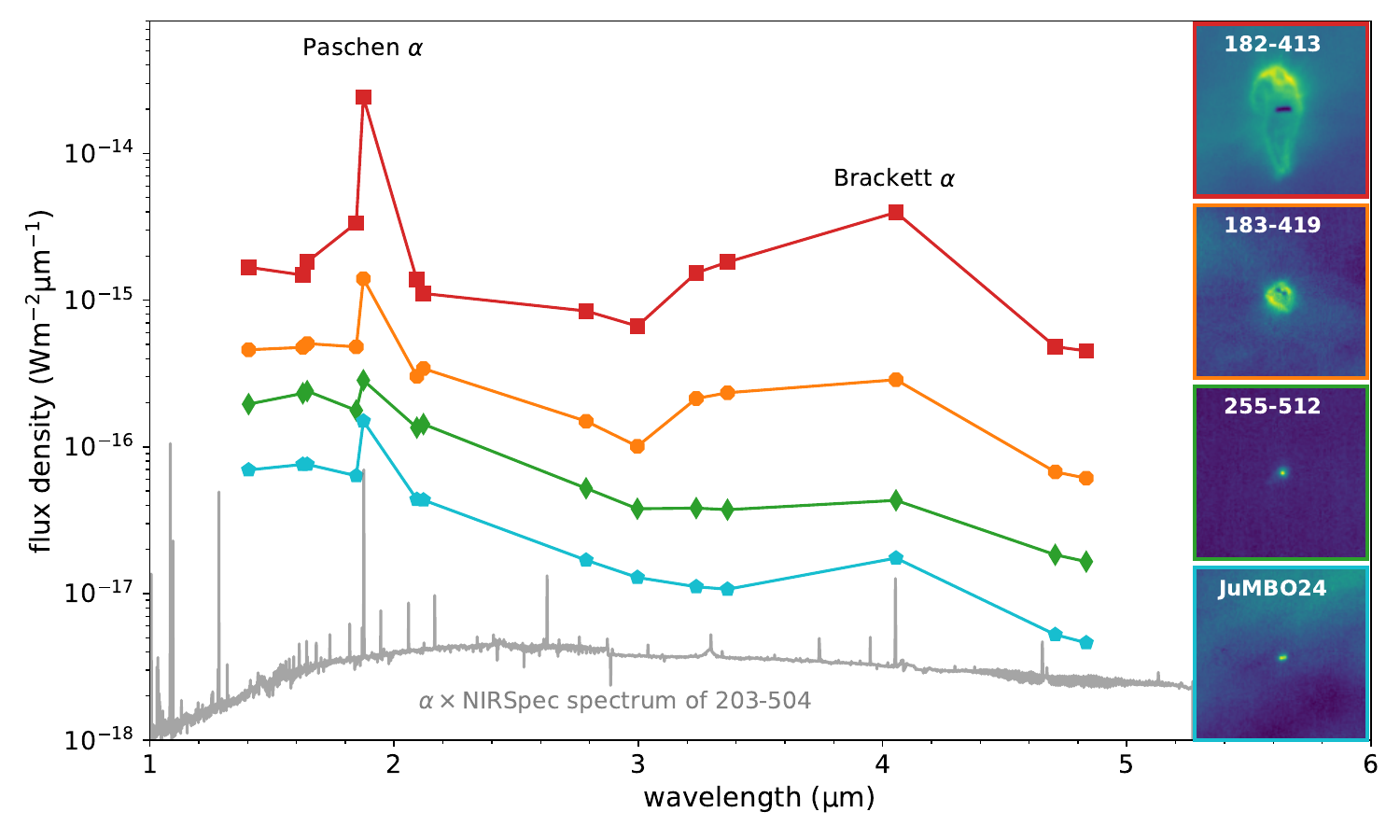}
    \caption{JWST-NIRcam SEDs from the \textit{PDRs4All} program of three Orion Proplyds (182-413, 183-419, and 255-512) and of JuMBO24, and their corresponding $4''\times4''$ image in the 
    F187N Pa $\alpha$ filter.  
    The order, from top to bottom, of the SEDs and images are the same. 
    In addition, the border of the images are in the colors of the corresponding SEDs.
    For clarity, the errors bars on the SEDs are not shown.
    The NIRSpec spectrum of 203-504, multiplied by a factor $\alpha=5\times10^{-4}$ to avoid overlaps with the SEDs, is shown in gray.
     }
    \label{fig:SED_proplyds_comp}
\end{figure*}

We relate the SEDs presented Fig.~\ref{fig:SED_proplyds} to the typology from Section~\ref{sect:typology}. 
Sources of {\it type I} and {\it II} systematically exhibit strong excess emission at 1.87~\textmu{}m and 4.05~\textmu{}m, in the Pa and Br $\alpha$ lines, respectively. These originate from the bright IF visible in the line emission maps (Fig.~\ref{fig:map_lines0}, Fig.~\ref{fig:Paschen_images}). 
The one exception is 191-232, where the flux at 1.87~\textmu{}m and 4.05~\textmu{}m were discarded.

Sources classified as {\it type III} have SEDs with much weaker or no excess at 1.87 and 4.05~\textmu{}m.
Instead, they have a blackbody-like shape with no strong features (see 201-534, 213-533, 218-529, and 239-334;  Fig.~\ref{fig:SED_proplyds}). 
However, 232-453, which is of {\it type III}, still has a SED with excess of emission in the Pa and Br $\alpha$ lines similarly to {\it type I} and {\it II} sources. 
Since no IF is observed for 232-453, this must originate from the spatially unresolved circumstellar medium, for example from accretion processes, which are known to produce strong hydrogen lines \citep[e.g.][]{rogers_determining_2024}, or from an unresolved IF.

The SED of 203-506 has a peculiar shape due to its morphology and to the flux extraction method. The edge-on disk obstructs most of the light emitted by the central star. 
However, the excess emission observed in the SED ([\ion{Fe}{ii}] line at 1.64~\textmu{}m and the H$_2$ line at 2.12~\textmu{}m) are consistent with the spatial features observed in the emission maps of the figures presented in Appendix~\ref{sect:All_maps}.

234-104, categorized as ``ambiguous'' Section~\ref{sect:typology}, has a blackbody-shaped SED similar to {\it type III} sources. 
The other ambiguous source, 180-218, has a SED that presents strong H$_2$ emission, possible water ice absorption at 3~\textmu{}m and possible PAHs emission at 3.3~\textmu{}m. Ignoring the 3.3~\textmu{}m flux, it is similar to that of 203-506.

We emphasize that the shape of the derived SEDs strongly depends on the size of the aperture used to extract the flux. Notably, the observed Pa and Br $\alpha$ excesses strongly depend on the quantity of IF contribution is evaluated. 
Thus, the link between the typology derived from the images and that derived from the SEDs must then be carefully established.

In Fig.~\ref{fig:SED_proplyds_comp}, we compare the SED of three sources categorized as {\it type I} or {\it II} (corresponding to proplyds, cf. Section~\ref{sect:typology}) with the NIRSpec spectrum obtained as part of the \textit{PDRs4All} program of 203-504, a well-studied proplyd \citep{schroetter_2025_504}.
This figure demonstrates that the spectral signatures visible in these SEDs are consistent with the emission lines in the proplyd 203-504, notably the strong Hydrogen recombination lines Pa and Br $\alpha$ at 1.87 and 4.05~\textmu{}m, respectively. In some cases, we also see the PAH features at 3.3~\textmu{}m originating from the disk and/or from the photoevaporative wind.

\subsection{SEDs of JuMBOs and and their comparison to ONC disk SEDs}

\begin{figure}[h]
     \includegraphics[width=\linewidth]{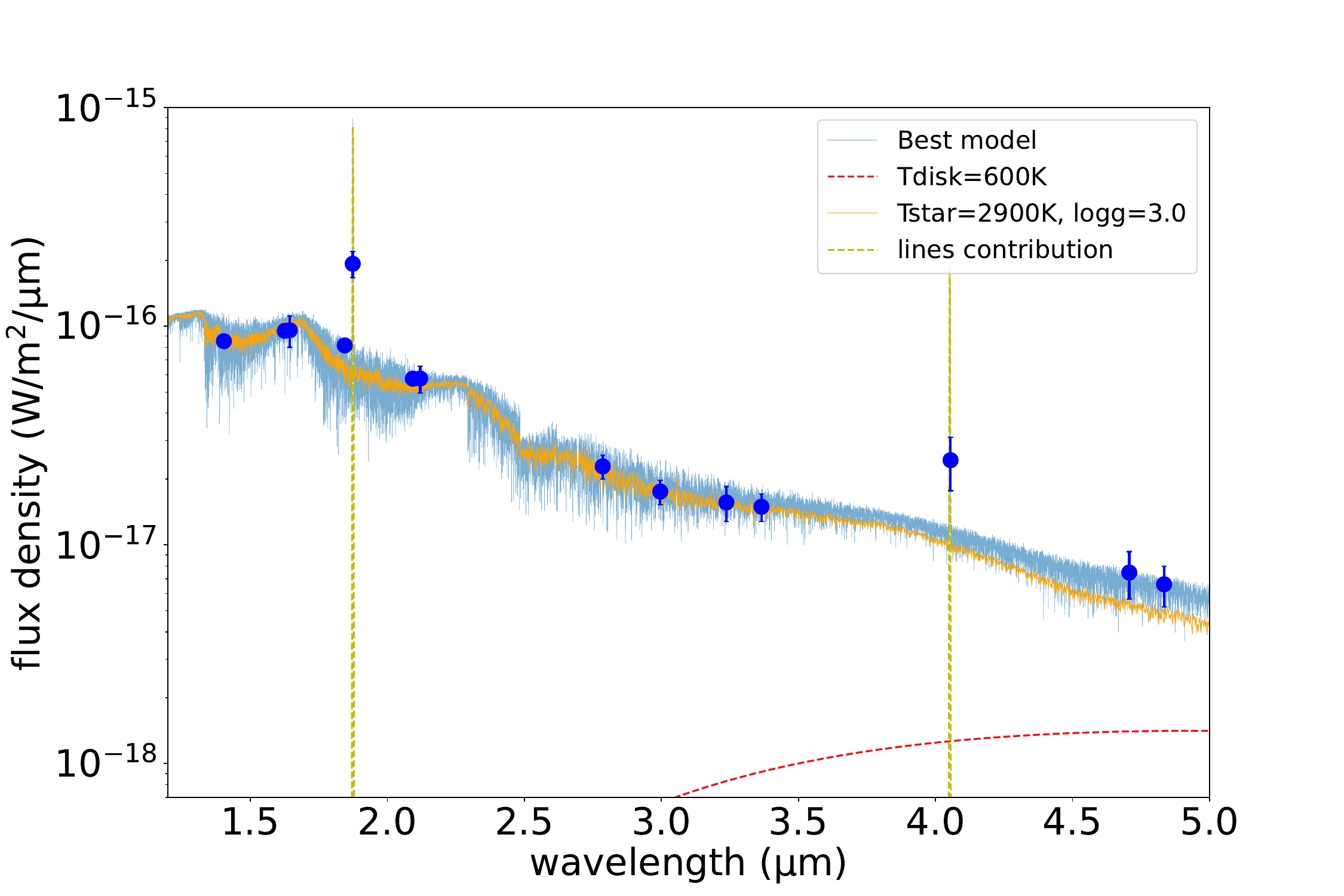}
     \caption{Best fit model of JuMBO 24 SED (in blue) obtained from the Phoenix model combined with a disk model. The dashed yellow lines are Gaussians with widths $\sigma=10^{-3}$~\textmu{}m, centered at 1.87~\textmu{}m and 4.05~\textmu{}m, respectively. The dashed red line corresponds to a blackbody at a temperature of $T_{\rm disk}=\Tdisk$K. The visual extinction is $A_{\rm V}=1.5$. The dashed orange lines corresponds to the spectrum of the central body given by the Phoenix model for $T_{\rm eff}=\Tstar$~K, $\log g$=\logg. The blue dots correspond to the JWST photometric points for JuMBO24.}
     \label{fig:JuMBO24_SED_fit}
\end{figure}

Currently, the nature of JuMBOs is not well constrained. For instance, \citet{luhman_candidates_2024} showed with spectroscopic data that some JuMBOs are, in fact, background reddened stars. 
Spectroscopic data will allow for a better assessment of the true nature of the JuMBO sample (McCaughrean et al., in prep). 

To contribute to this assessment, we compare the \textit{PDRs4All} SEDs of JuMBOs to those of ONC disks.
The SEDs of JuMBOs are presented in Fig.~\ref{fig:SED_proplyds}. 
Among the sample of 11 JuMBOs, the primary source is detected in all filters for only five systems. 
For JuMBO26, 27, 28 and 29 only the primary sources are bright enough to build the SED (see Fig.~\ref{fig:kaleidoscope_JuMBOs}).
For JuMBO24, both objects are bright, but are merged in the images at wavelengths above 2.3~\textmu{}m. Thus, the SED shown in Fig.~\ref{fig:SED_proplyds} merges that of the two objects. In Appendix~\ref{sect:Appendix_JuMBO24} we derive the SED for JuMBO24a and JuMBO24b in the short wavelength filters of NIRCam and show that they are similar. 

The SED of JuMBO26, 27, 28 and 29 have a blackbody shape and peak around 1.8-2~\textmu{}m. They do not present any features, similarly to the SEDs of {\it type III} sources.
Conversely, JuMBO24's SED shows strong excess emission at 1.87 and 4.05~\textmu{}m, following the SED shape of {\it type I} and {\it II} sources. 
In Fig.~\ref{fig:SED_proplyds_comp}, we compare JuMBO24's SED with that of three {\it type I-II} sources as well as with the NIRSpec spectrum of the proplyd 203-504. 
The similarities between the SEDs suggest that JuMBO24 may be of the same nature as {\it type I-II} sources, but smaller in size. 
Note that JuMBO24's SED is also similar to that of 232-453, neither source presenting extended Pa $\alpha$ emission. As we discussed, the Pa $\alpha$ emission in 232-453 is expected to originate from an unresolved IF or from ongoing accretion in the presence of a circumstellar disk. 

To assess the viability of this hypothesis, we fit the SED of JuMBO24 with a basic model that includes a spectrum of a young star with an irradiated protoplanetary disk.
In this analysis, we considered JuMBO24 as a single sources (rather than a close binary), simply dividing its measured SED by two (cf. Appendix~\ref{sect:Appendix_JuMBO24}).
 To model the emission of the young star's photosphere, we used the spectra of the \textsc{Phoenix} model \citep{hauschildt_phoenix_1999,husser_phoenix_2013}, whose parameters are the effective temperature of the central body, $T_{\text{eff}}$, the surface gravity, $\log g$, and the metallicity. We assumed a solar metallicity and $\log g=3.0$.
The thermal emission of the disk was modeled with a blackbody at the temperature $T_{\text{disk}}$. {Although a simple blackbody is not physically accurate, it has the advantage of limiting the number of free parameters. Furthermore, more sophisticated disk models consider an isolated disk heated by the central star, which may not be appropriate. Indeed, the disks in ONC are externally heated by the massive stars from the Trapezium, which produces out of equilibrium emission in the NIR and MIR range (PAHs, very small grains, etc).}
The hydrogen recombination lines are considered as two Gaussians, centered on the Pa and Br $\alpha$ wavelengths at 1.87~\textmu{}m and 4.05~\textmu{}m, respectively, with a width $\sigma=0.01$~\textmu{}m. 
Lastly, the effect of dust extinction on the line of sight, parameterized by the visual extinction $A_{\rm V}$, is considered using the extinction curve of the model of \citet{Weingartner_Draine_2001} with $R_\text{V} = 5.5$ {\citep[e.g.,][]{Cardelli_1989, Fang_2021}}.

The best fit model that minimizes the reduced $\chi^2$ to the data yields $T_{\text{eff}} = \Tstar$~K, $T_{\text{disk}} = \Tdisk$~K and A$_{\rm V}$ = \Av, and is shown in Fig.~\ref{fig:JuMBO24_SED_fit}.
This confirms that the SED of JuMBO24 is compatible with that of a binary system of low-mass stars with au-scale circumstellar irradiated disks.
In the F187N filter, the binary system is resolved and no extended emission encompassing both sources is seen, which is why we exclude the possibility of a single large IF with a circumbinary disk. 

\citet{rodriguez_radio_2024} reported radio emission from JuMBO24 in archival Karl G. Jansky Very Large Array observations.
In Appendix~\ref{sect:EMradio}, we assume that this originates from the free-free emission of electrons at an IF and derive the associated emission measure, $EM_{\rm radio} = (9.0\pm2.0)\:10^{5} {\rm cm}^{-6}\  {\rm pc}$. We also convert the Pa $\alpha$ emission into $EM_{\rm Pa \alpha}$, still assuming emission from an IF, and find  
$EM_{\rm Pa\alpha} = 5.8 \times 10^{6}$ cm$^{-6}$ pc. Both EMs have the same order of magnitude, suggesting that the observed radio emission can be explained by the presence of an IF.

\section{NIR properties of ONC disks}\label{sect:NIRproperties}

We investigated the properties of the ONC disks as seen in the NIRCam images. We started with the disk and radius of the IF, and then turned to the Pa $\alpha$ intensity.

\subsection{Ionization front and disk sizes}\label{sect:disk_IF_size}

\subsubsection{Method}

\begin{figure*}
    \centering
    \includegraphics[width=\linewidth]{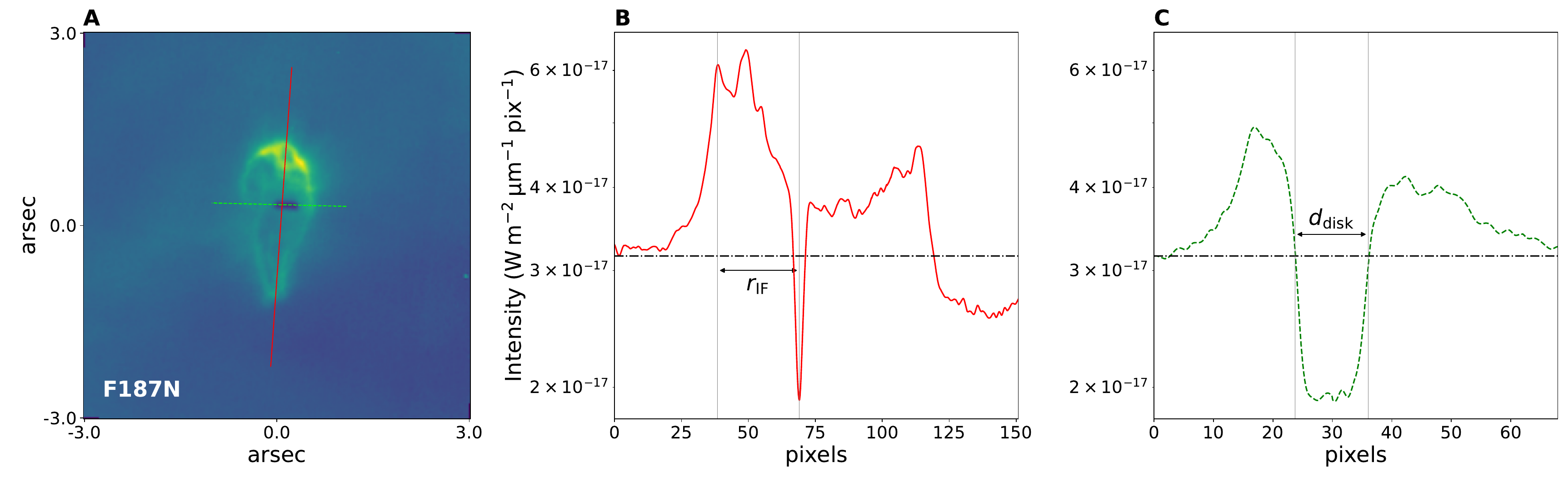}
\caption{(\textbf{A}): NIRCam image of 182-413 in the F187N filter with 6''$\times$6'' FOV. 
(\textbf{B}): intensity profile extracted from the cut shown with the continuous red line in panel \textbf{A}, along the IF. The horizontal dash-dotted black line corresponds to the median of intensity of the cropped image.
(\textbf{C}): intensity profile extracted along the cut shown with the dashed green line in panel \textbf{A}. The limits of the IF are shown with the vertical dashed lines.}
    \label{fig:profile_disk_cocoon}
\end{figure*}

Fig.~\ref{fig:profile_disk_cocoon} summarizes the process used to measure the disk and IF radius.
The IF radius ($r_{\rm IF}$) is defined as the distance between the central star and the tip of the IF. 
It is estimated using the intensity profile of the source extracted from a cut along the IF length in the F187N filter, as shown in Fig.~\ref{fig:profile_disk_cocoon} (\textbf{A}) (continuous red line). 
We locate the star at the central peak of intensity of the profile. 
In the case where the disk is edge-on and obstructs the star, we consider the peak of absorption to locate the star, as shown in panel (\textbf{B}).
If the central star is saturated, we locate it at the center of the saturated pixels.
The tip of the IF is placed at the peak of intensity in the profile that is closest to the ionizing source, following \citet{aru_kaleidoscope_2024}. 
The IF radius is then derived as the distance between the star and the tip of the IF, with an uncertainty of $\pm$1 pixel.
For the disk, the absorption profile is extracted from a cut  along the disk major axis (dashed green line panel in \textbf{A}) in the F187N filter.
The outer edges of the disk are then determined using the intersection between the line profile (continuous green line in the panel \textbf{C}) and the median of the local background intensity $\pm$20\% (dashed line in the panel \textbf{C}), depending on the surrounding nebular emission. 
The distance between the two intersection points is then considered to be a direct measure of the disk diameter. 
We estimate the error from $\pm$10--20\% around the median, which gives a general error of $\pm$1 pixel.

As the disks are observed in silhouette, the F187N filter's point spread function (PSF) tends to reduce the observed size of the disks. Thus, the measured values are corrected from the PSF half-width at half-maximum (HWHM) considering
\begin{equation}\label{eq:deconvolution}
r_{\rm disk} = \sqrt{r_{\rm obs}^2 + {\rm HWHM}_{\rm F187N}^2}\ ,
\end{equation}
where $r_{\rm disk}$ corresponds to the estimated disk radius corrected for the PSF, $r_{\rm obs}$ is the observed radius, and ${\rm HWHM}_{\rm F187N}$ is the HWHM of the F187N filter's PSF.

From the sample presented in Section~\ref{sect:sample}, we measure the disk radius for 22 sources and the IF radius for 23. For the IF, we only consider the sources for which the tip of the IF is clearly distinct from the central star.
The obtained disk and IF radii are given in Table~\ref{tab:coordinates}.

\subsubsection{Disk radius}\label{sect:disk_radius}

\begin{figure*}[h]
    \centering
    \includegraphics[width=\linewidth]{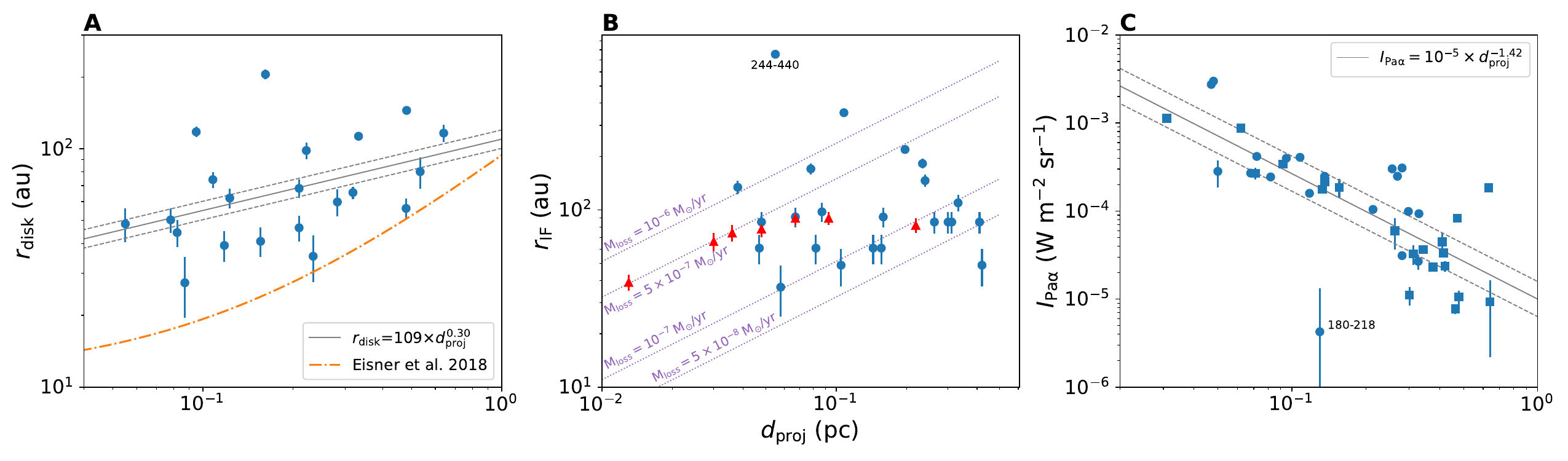}
    \caption{Disk radius (\textbf{A}), IF radius (\textbf{B}), and Pa $\alpha$ intensity (\textbf{C}) as a function of the projected distance ($d_{\rm proj}$) to the ionizing source. The solid black lines correspond to the linear regression. 
    The dashed lines mark the regression at $\pm1\sigma$. The dash-dotted line in panel \textbf{A} corresponds to the correlation obtained with ALMA observation at 850~\textmu{}m by \citet{eisner_protoplanetary_2018}. 
    The red triangles in panel \textbf{B} correspond to the IF radii measured by \citet{aru_kaleidoscope_2024}. 
    The purples lines are the expectation from the model of \citet{johnstone_1998} for given mass-loss rates.}
    \label{fig:measure_vs_dist}
\end{figure*}

\begin{figure}[h]
    \centering
        \includegraphics[width=0.85\linewidth]{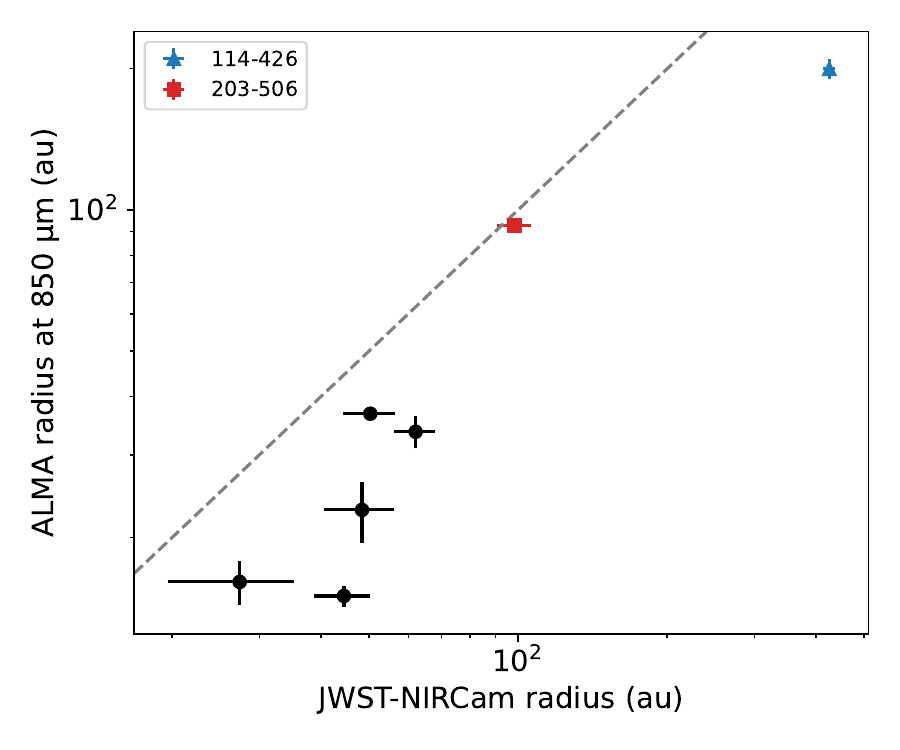}
    \caption{{Disk radii measured with ALMA at 850~\textmu{}m (from literature) as a function of the measured radii in NIRCam-F187N filter (this work).
    The ALMA radius for 114-246 and 203-506 are from \citet{bally_2015_114-426, berne_far-ultravioletdriven_2024}, respectively. The remaining ALMA values are from \citet{eisner_protoplanetary_2018}.} } 
    \label{fig:radius_ALMA_JWST}
    
    \includegraphics[width=0.8\linewidth]{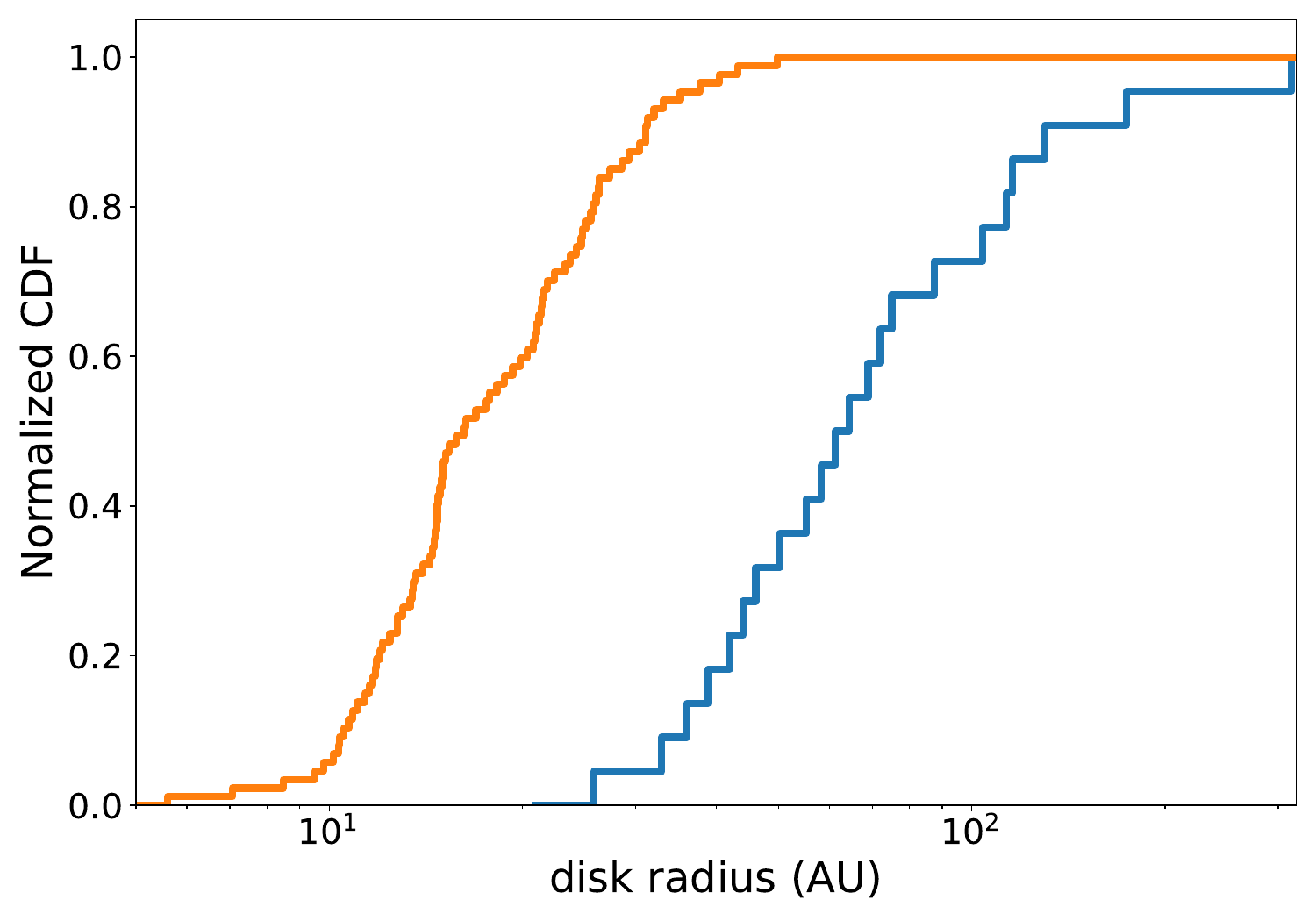}
    \caption{Cumulative distribution function (CDF) of disk radii in the ONC from various studies. Blue: this work. Orange: CDF from \citet{eisner_protoplanetary_2018} obtained from ALMA observations of dust continuum emission.} 
    \label{fig:CDF_radius}
\end{figure}

Fig.~\ref{fig:measure_vs_dist}\textbf{A} shows the disk radii as a function of the projected distance to the main ionizing source, which is $\theta^1$ Ori C except for 244-440 where it is $\theta^2$ Ori A. 114-426 is excluded from this analysis because of its much larger size.
The disk radius increases with the projected distance to the ionizing source. 
The regression provides a relationship of the form $r_{\rm disk}=109\times d^{0.30}$, with $r_{\rm disk}$ in astronomical units and $d$ the projected distance in parsecs.
To assess the significance of this correlation, we randomly sampled the disk radii on their corresponding uncertainty ranges. We then derived the associated Pearson coefficient, $R$. This process was repeated $10^4$ times. We finally considered the median value of $R$, with its standard deviation as the uncertainty.
This provides a Pearson coefficient of $R=0.41\pm0.03$.
This moderate correlation is affected by multiple factors, such as the projection effect along the line of sight, different initial disk conditions and evolutionary pathways, or high proper motion.

A linear trend between the disk radius and projected distance has already been observed for the ONC, notably with the Atacama large millimeter array (ALMA) in dust continuum emission at 850~\textmu{}m on a sample of 104 disks \citep[][cf. Fig.~\ref{fig:measure_vs_dist}\textbf{A}]{eisner_protoplanetary_2018}.
To follow up on the comparison of disk radii as observed with JWST and ALMA, we present in Fig.~\ref{fig:radius_ALMA_JWST} {the disk radii obtained from the NIRCam images and their associated radii obtained from ALMA observations at 850~\textmu{}m. The ALMA values of 203-506 and 114-426 are from \citet{berne_far-ultravioletdriven_2024, bally_2015_114-426}, respectively. The other ALMA values are from \citet{eisner_protoplanetary_2018}. The NIRCam derived disk radii are systematically larger than the ALMA derived disk radii for these sources.}
Fig.~\ref{fig:CDF_radius} shows the cumulative distribution function of disk radii measured from the NIRCam images as well as those obtained by \citet{eisner_protoplanetary_2018} from ALMA data. 
The CDF obtained with NIRCam images indicates a mean radius for the disks of $\approx$50~au whereas the mean radius with ALMA is $\approx$25~au. Statistically, disks thus appear two times smaller when observed at submillimeter wavelengths than at near-IR wavelengths. This is also illustrated Fig.~\ref{fig:measure_vs_dist}\textbf{A}, where the obtained relation provides systematically smaller disks.
\citet{vicente_size_2005} estimated the radius of 149 disks in the ONC with HST's Wide Field Channel of the Advanced Camera for Surveys, in the visible. Their CDF follows closely that of the NIRCam CDF. In summary, disk as seen by JWST appear the same size when seen by HST, but are twice as small when seen by ALMA. 
This can be interpreted as evidence of dust radial segregation in disks. 
The pebble-sized grains, traced by the continuum at 850~\textmu{}m with ALMA, are concentrated in the innermost part of the disks because of the faster radial drift of large grains \citep[see review by][]{birnstiel2024_review}. The micrometer-sized grains that are seen in absorption in the near-IR by NIRCam remain more spatially extended \citep[e.g.,][]{Andrews_2020}.
Evidence of this process has already been observed in some sources in the ONC, such as 114-426 \citep{bally_2015_114-426, Ballering_2025_114-426} with NIRCam or 197-427 \citep{mann_submillimeter_2010} in Sub-Millimeter Array (SMA) data.

\subsubsection{Ionization front radius}\label{sect:IF}

Fig.~\ref{fig:measure_vs_dist}\textbf{B} shows the IF radii as a function of the distance to the main ionizing source. 
No correlation is observed. 
Performing the same analysis as Section~\ref{sect:disk_radius}, we obtain a Pearson coefficient of $R=0.06\pm0.04$.
This flat correlation between the IF radius and projected distance is also observed and discussed in Gupta et al. (in prep.).

We compare the empirical IF radii with the theoretical expectations for four different mass-loss rates, using the equation presented in \citet{winter_external_2022}: 
\[
 r_{\rm IF} = 1200 \times  \left(\dfrac{\rm M_{ loss}}{\rm 10^{-8}~M_{\odot}\: yr^{-1}} \right)^{2/3} \left(\dfrac{\Phi_{\rm EUV}}{10^{45}~ \rm s^{-1}}\right)^{-1/3} \left(\dfrac{d}{1~\rm pc} \right)^{2/3} ~ \rm au\ ,
\]
where $M_{\rm loss}$ is the mass-loss rate due to the external photoevaporation, $\Phi$ is the Lyman continuum emission of the ionizing source, and $d$ is the projected distance to the ionizing source. 
In Appendix \ref{sect:thetaOriC}, we derive $\Phi_{\rm EUV}=1.29\times10^{49}~\rm s^{-1}$ for $\theta^{1}$ Ori C.
Around 86\% of the IF radii align with a theoretical mass-loss rate in the range $\rm M_{loss}\in [5\times10^{-8}, ~10^{-6}]~M_\odot~yr^{-1}$. This is of the order of the typical values derived for the ONC \citep{johnstone_1998, henney1998_proplyds, henney1999_keckONC}.

The IF radii are compared with the radii measured by \citet{aru_kaleidoscope_2024} in H$\alpha$ emission with Very Large Telescope-MUSE observations (red triangles). We excluded data points for which only a lower limit is provided. 
We see from Fig.~\ref{fig:measure_vs_dist}\textbf{B} that the observed correlation IF radius-projected distance in \citet{aru_kaleidoscope_2024} disappears when a larger sample is considered.

\subsubsection{Connecting disk and ionization front radius to $G_0$}

\begin{figure}
    \centering
    \includegraphics[width=0.9\linewidth]{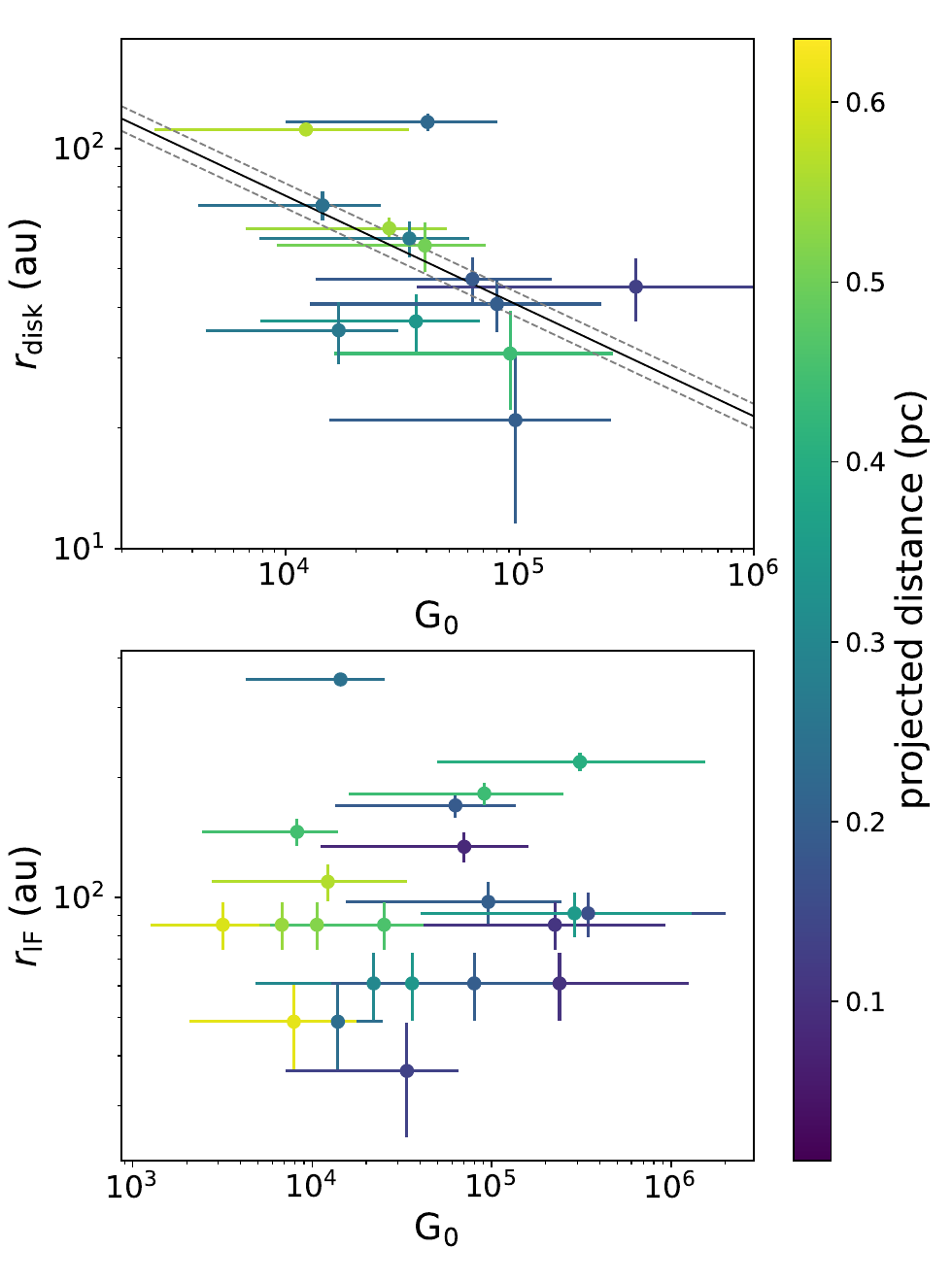}
    \caption{Disk and IF radii as a function of the incident FUV radiation field in unit of Habing as provided by \citet{anania_2025_G0}. The color gradient indicates the projected distance to the ionizing source.}
    \label{fig:measures_G0}
\end{figure}

 \citet{anania_2025_G0} estimated the incident FUV radiation flux, $ G_0$, expressed in units of the Habing flux\footnote{the Habing flux is $1.6\times10^{-3} ~\rm erg~s^{-1}~cm^{-2}$ } \citep{habing1968}, for disks in the ONC. 
Their method makes use of a density distribution to infer the 3D geometry from the 2D geometry, assuming spatial isotropy in the cluster and using \textit{Gaia} DR3 positions \citep{gaia2023_dr3}.
This provides an estimate of the true separation between the sources and the ionizing sources, which gives an estimate of the true incident FUV field. 

Fig.~\ref{fig:measures_G0} presents the disk and IF radii as a function of $G_0$ as provided by \citet{anania_2025_G0}.
We observe decreasing disk radii for increasing $G_0$, with a Pearson coefficient of $R=-0.40\pm 0.14$.
This is compatible with the correlation between the disk radii and projected distance to the ionizing source (cf. Fig.~\ref{fig:measure_vs_dist}\textbf{A}). 
This confirms that disks are truncated more rapidly when exposed to a stronger FUV radiation field, suggesting that the mass-loss rate increases when the intensity of the FUV radiation field increases. 
Conversely, no correlation between the IF radii and $G_0$ is observed ($R=0.11\pm0.07$), whereas a decreasing IF radius with increasing $G_0$ is expected by theoretical models \citep{johnstone_1998} as long as a constant photoevaporative flow in the PDR is assumed.

\subsection{Paschen $\alpha$ intensity}

We derived the Pa $\alpha$ intensity for all the sources that are not saturated in the F187N and F182M filters, combining the NIRCam images from the GTO and \textit{PDRs4All} program datasets. 
We used the equation from \citet{Chown2025_calibration} (cf. Equation~\ref{eq:recipes_lines}) and the flux extracted in Section~\ref{sect:SED} to derive the Pa $\alpha$ intensity.

Fig.~\ref{fig:measure_vs_dist}\textbf{C} shows the Pa $\alpha$ intensity ($I_{\rm Pa\alpha}$) of the ONC disks as a function of the projected distance to the ionizing source.
We observe a decrease in the Pa $\alpha$ intensity with increasing projected distance to the ionizing source. 
The associated Pearson coefficient is $R = 0.77\pm0.02$. 
180-218 is a clear outsider, with a Pa $\alpha$ intensity $\sim 10^2$ times lower than expected based on the observed trend. Its SED (cf. Fig~\ref{fig:SED_proplyds}) indicates a flux at 1.87~\textmu{}m lower than the continuum, suggesting an over subtraction of the background. 
The Pa $\alpha$ image Fig.~\ref{fig:Paschen_images} and the F187N image Fig.~\ref{fig:HC182} of 180-218 show a bright structure close to the source, which contributes to the estimated background flux and leads to this over-subtraction.

\section{Linking ONC disk properties with PDRs}\label{sect:ONCdisk_PDR}

\begin{figure}
    \centering
    \includegraphics[width=\linewidth]{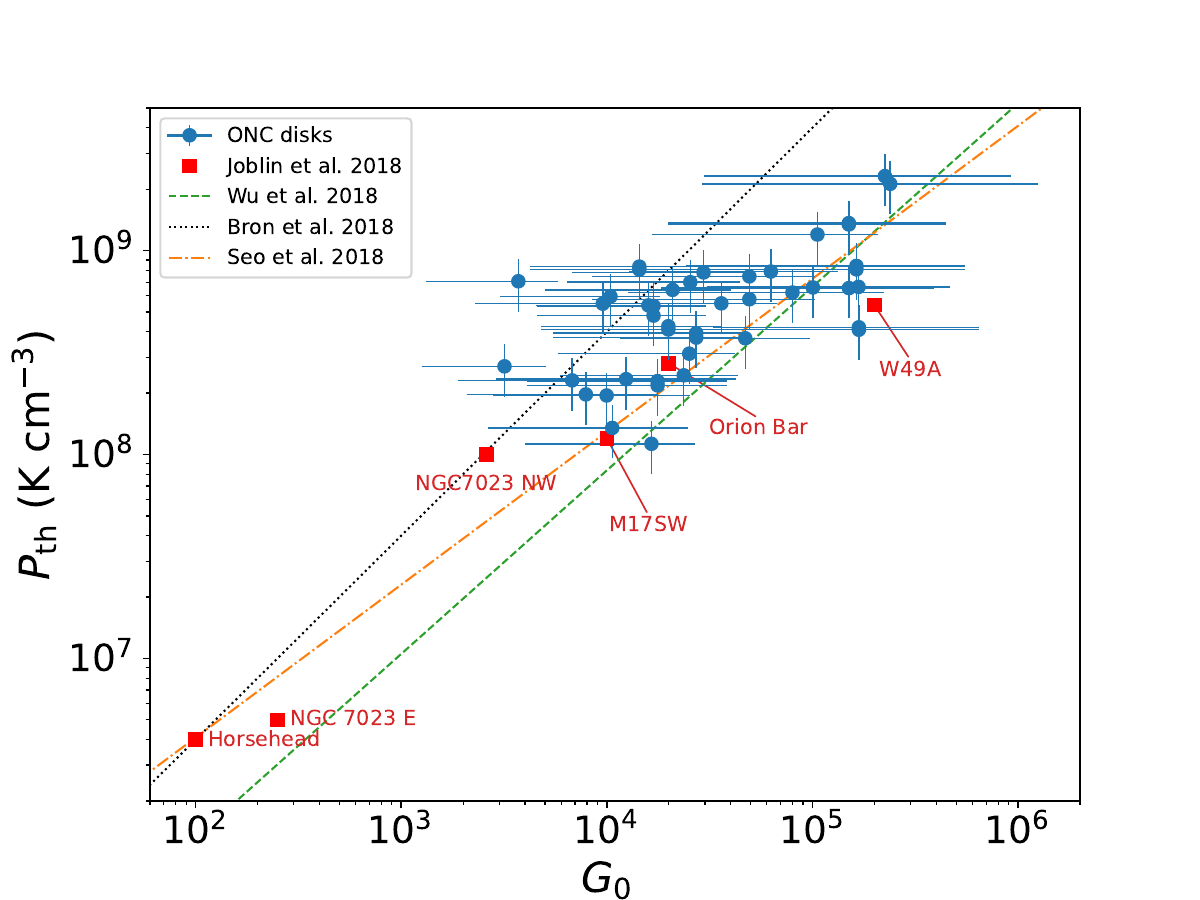}
    \caption{Thermal pressure in the PDR of ONC disk as a function of the incident FUV radiation field $G_0$. The uncertainties on $P_{\rm th}$ arise from those associated with the length of the emitting layer, considered as the standard deviation of the IF radii, and from the uncertainty on the temperature of the HII region. The red squares are values for various PDRs reported by \citet{joblin_2018_PDR_hershell} and references therein. The dotted lines correspond to the relations obtained in the models of \citet{Bron2018_PDR}, considering an O-star for the irradiating source. The dashed black line corresponds to the empirical relation obtained in \citet{Wu_2018_PDR} for Trumpler 14 in the Carina Nebula. The dash-dotted orange line correspond to the analytical relation of \citet{Seo2019_PDR_Tr14} \citep[also discussed in][]{Wolfire2022_PDR}, using the values of $\theta^1$ Ori C derived in Appendix \ref{sect:thetaOriC}. 
    }
    \label{fig:Pth}
\end{figure}

The external photoevaporation of protoplanetary disks is directly related to the interaction of FUV photons with the gas, forming a PDR at the disk surface. Early models of external photoevaporation were based on PDR physics \citep{johnstone_1998, storzer_photodissociation_1999, adams_photoevaporation_2004}. 
However, it has been difficult to constrain such models with observations because the cooling of PDRs occurs mainly in the near- to far-IR. 
Observatories such as ISO, Spitzer, and Herschel, providing an angular resolution of well above 1'' for spectroscopy, were unable to detect characteristic PDR lines against the bright background in Orion (except in some cases : \citealt{champion_2017_hershell,  vicente_2013_HST10}). 
However, studies of PDRs in the ISM benefited greatly from these space missions. 
In particular, observational results suggest that the thermal pressure ($P_{\rm th}$) in interstellar PDRs is directly related to $G_0$, with increasing $P_{\rm th}$ for increasing $G_0$ (\citealt{joblin_2018_PDR_hershell, Wu_2018_PDR, Bron2018_PDR, Seo2019_PDR_Tr14} and Fig.~\ref{fig:Pth}).
{The competition of $G_0$ against the gas density was discussed in the case of photoevaporating PDRs by \citet{Maillard2021}, showing that some configurations could lead to IF-DF merging. In this section, we attempt to identify whether the  $P_{\rm th}$-$G_0$ correlation extends to externally irradiated disks.}

To derive the pressure in the PDRs of the ONC disks, we consider mass loss by photoevaporation from the disk and conservation of mass at the IF, where the gas flows outward into the HII region. We use the continuity equation at the IF: $n_{\rm I}c_{\rm I}=n_{\rm II}c_{\rm II}$, 
where $n_{\rm I}$, $c_{\rm I}$ are the hydrogen number density and sound speed within the PDR, respectively, and $n_{\rm II}$, $c_{\rm II}$ are the same parameters for the \ion{H}{ii} region.
Since $c \propto T^{1/2}$, this yields $n_{\rm I}=n_{\rm II}\left({{T_{\rm II}}/{T_{\rm I}}} \right)^{1/2}.$ The thermal pressure within the PDR is $P_{\rm th}=n_{\rm I}T_{\rm I}$; thus, $P_{\rm th} = n_{\rm II}({T_{\rm I}}{T_{\rm II}})^{1/2}$.
We relate the hydrogen density at IF in the \ion{H}{ii} region $n_{\rm II}$ to the emission measure as $EM=n_{\rm II}^2L$ with $EM$ in $\rm cm^{-6}~pc$ and $L$ the length of the emitting layer in pc. This gives
\begin{equation*}
    P_{\rm th} = \left(\dfrac{EM~{T_{\rm I}}~{T_{\rm II}}}{L}\right)^{1/2}.
\end{equation*}
$EM$ scales with the Pa $\alpha$ intensity \citep{balog_photoevaporation_2008} as $I_{\rm Pa\alpha} = 2.51\times10^{-19}EM$, with $I_{\rm Pa\alpha}$ in $\rm erg~s^{-1}~cm^{-2}~arsec^{-2}$ and $EM$ in $\rm cm^{-6}~pc$. 
We thus derive $EM$ for all the non-saturated sources in the \textit{PDRs4All} and GTO data. The temperature within the PDR has been constrained by observations and PDR-models to $T_{\rm I}\approx1000$~K \citep{champion_2017_hershell, berne_far-ultravioletdriven_2024}. 
\citet{boyden2025_radio_lines} derived electron temperatures at the IF of ONC disks with ALMA and Submillimeter Array observations. They obtained values between 6000 and 11000~K. This interval is considered for $T_{\rm II}$. For the length of the emitting layer, $L$, we assume the median of the IF radii with an uncertainty given by the standard deviation of the measures, obtaining $L=100\pm50$~au.

Fig.~\ref{fig:Pth} shows the derived $P_{\rm th}$ as a function of $G_0$ for ONC disks. The values are in the range of $P_{\rm th} \sim 10^{8-9}~\rm K~cm^{-3}$. This is lower than the values derived from the analysis of the H$_{2}$ ro-vibrational emission lines observed with JWST in 203-506 and 203-504 ($P_{\rm th} \in [7\times10^{8}-10^{10}]~\rm K~cm^{-3}$ and $P_{\rm th} = 3.8\times10^{10}~\rm K~cm^{-3}$, respectively (see \citealt{berne_far-ultravioletdriven_2024, schroetter_2025_504}). 
These pressures correspond to the DF situated at the disk surface (see Section \ref{sect:typology}), whereas the pressures shown in Fig.~\ref{fig:Pth} correspond to that of the neutral atomic gas at the IF, thus downstream in the photoevaporation flow, where the pressure is expected to be lower. 
Fig.~\ref{fig:Pth} also shows that there is a correlation 
between $P_{\rm th}$ and  $G_0$ for ONC disks (correlation coefficient $R=0.54\pm0.06$), as observed in 
interstellar PDRs. 
\citet{joblin_2018_PDR_hershell}
suggest that the relationship between $P_{\rm th}$ and  $G_0$  
can be approximated by a linear relation  $P_{\rm th} \propto G_0$. Similarly, using Hershel data from Trumpler 14, \citet{Wu_2018_PDR} find $P_{\rm th} \propto G_0^{0.9}$ (shown in Fig.~\ref{fig:Pth}).
Using dynamical 1D PDR models, \citet{Bron2018_PDR} show that compression by the dissociation or IF in an interstellar PDR produces a $P_{\rm th} \propto G_0$ relation, which we report in Fig.~\ref{fig:Pth}.  \citet{Seo2019_PDR_Tr14} consider a photoevaporating clump and analytically show that $P_{\rm th} \propto G_0^{0.75}$ is predicted (also shown in Fig.~\ref{fig:Pth}). 
ONC disk points in Fig.~\ref{fig:Pth} appear to follow a similar trend of a power law relation with $G_0$, and visually agree well with the law discussed in \citet{Seo2019_PDR_Tr14}. This suggests that the pressure in the wind is determined by mechanisms similar to those described for interstellar PDRs. However, the best fit to ONC disk points in Fig.~\ref{fig:Pth} yields $P_{\rm th} \propto G_0^{0.26\pm0.03}$, which is somewhat flatter than the trends reported in the references discussed above. This suggests that while $P_{\rm th}$ increases with $G_0$ in the winds of ONC disks as in ISM PDRs, the detailed physics and dynamics likely differ between these two environments. This demonstrates the interest in developing models based on state of the art PDR physics and chemistry that also include disk and wind dynamics in a self-consistent manner \citep[see discussion in][]{planet_formation_collab_2025}.

\section{Conclusion}

Using the  NIRCam images from the \textit{PDRs4All} ERS 1288 program, we derived the [\ion{Fe}{ii}], Paschen $\alpha$, H$_2$ 1-0 S(1), and PAH  emission map for 22 protoplanetary disks. 
We used these maps to define three types of disks, depending on the position of the IF with respect to the disk surface (cf. Fig.\ref{fig:scheme_category}):  
{\it type I} are the sources {where the IF and disk surface are not separated in NIRCam images (i.e., are separated by $\lesssim$ 25 au)}, {\it type II} are those where the IF and disk surface are well separated ({by $\gtrsim$ 25 au}), and {\it type III} are the sources with no IF.
Within {a projected distance of} 0.2 pc from $\theta^1$ Ori C, $90~\%$ of the sources are of {\it type I} or {\it II}. Above 0.2 pc, this trend is inverted, with $70~\%$ of the sources being of {\it type III}.
From a morphological point of view, {\it type I} and {\it type II} sources are consistent with the EUV-dominated flows and FUV-dominated flows described in \citep{johnstone_1998}, respectively. 
However, we do not find any evidence of a clear separation between the two categories at a distance of 0.03 pc to $\theta^1$ Ori C, as was suggested by \citet{johnstone_1998}.

We derived the SED of 85 ONC disks and 11 JuMBOs from the \textit{PDRs4All} images. Discarding SEDs with saturation or non-detection, we analyzed the SEDs of 23 ONC disks and five JuMBOs. 
The spectral features are related to typology: the SED of {\it type I} and {\it II} sources exhibit strong excess emission in the Pa and Br $\alpha$ filters. Their spectral features are consistent with the NIRSpec spectrum of the well-studied proplyd 203-504.
Conversely, the SEDs of {\it type III} sources are mostly blackbody-like and do not exhibit any strong features. 
All of the JuMBOs but JuMBO24 have SEDs that are similar to those of {\it type III} sources. Instead, the SED of JuMBO24 shows strong similarities with that of {\it type I} and {\it II} sources. JuMBO24 SED is successfully reproduced with a model consisting of a low-mass binary star with a circumstellar irradiated disks, suggesting that JuMBO24 may be a proplyd binary.

Combining the NIRCam images from GTO 1256 and the \textit{PDRs4All} 1288 programs, we measured the disk radius for 22 sources and the IF radius for 23 sources.
The measurements obtained for the disks radii are, on average, double the size of those obtained with ALMA at submillimeter wavelengths, representing the dust continuum emission.
This is interpreted as evidence of dust radial segregation within the disk: the millimeter-sized grains are concentrated close to the central star because of rapid radial drift.

We observe increasing disk radii with projected distance to $\theta^1$ Ori C $d_{\rm proj}$, following $r_{\rm disk}\propto d_{\rm proj}^{0.30}$, and with increasing $G_0$. This confirms that disks are truncated more rapidly when exposed to a stronger FUV radiation field. 
However, no correlation between IF radius and {either projected distance to $\theta^1$ Ori C} or $G_0$ is observed.
Comparison of the IF radii with theoretical expectations shows that the mass-loss rates for the sample are in the range $\rm M_{loss}\in [5\times10^{-8}, ~10^{-6}]~M_\odot~yr^{-1}$.

We derived the Pa $\alpha$ intensity for all the non-saturated sources contained in the \textit{PDRs4All} and GTO 1256 programs.
The Pa $\alpha$ intensity and the IF radius were used to derive the thermal pressure ($P_{\rm th}$) within the PDRs associated with the ONC disks.
We find that $P_{\rm th}$ increases with $G_0$
as is observed in PDRs of the ISM. However, our results suggest that the exponent in the power-law relationship between $P_{\rm th}$ and $G_0$ is lower for ONC disks ($\sim 0.26$) than for galactic PDRs ($\sim 0.75-1$), pointing to differences in the PDR physics involved in ONC disk winds with respect to galactic PDRs.

\section{Data availability}

Table \ref{tab:coordinates} is available at the CDS via \url{https://cdsarc.cds.unistra.fr/viz-bin/cat/J/A+A/708/A111}.

\begin{acknowledgements}
The data were obtained from the Mikulski Archive for Space Telescopes at the Space Telescope Science Institute, which is operated by the Association of Universities for Research in Astronomy, Inc., under NASA contract NAS 5-03127 for JWST. These observations are associated with programs \#1288.
OB, IS are funded by the Centre National d'Etudes Spatiales (CNES) through the APR program. 
This research received funding from the program ANR-22-EXOR-0001 Origins of the Institut National des Sciences de l’Univers, CNRS. 
  Part of this work was supported by \emph{ESO}, project
  number Ts~17/2--1.
JRG thanks the Spanish MCINN for funding support under grants PID2023-146667NB-I00.
  C.B. is grateful for an appointment at NASA Ames Research Center through the San Jos\'e State University Research Foundation (80NSSC22M0107). C.B. acknowledges support from the Internal Scientist Funding Model (ISFM) Laboratory Astrophysics Directed Work Package at NASA Ames.
  TJH acknowledges UKRI guaranteed funding for a Horizon Europe ERC consolidator grant (EP/Y024710/1) and a Royal Society Dorothy Hodgkin Fellowship. 
   This project is co-funded by the European Union (ERC, SUL4LIFE, grant
agreement No 101096293). AF also thanks project PID2022-137980NB-I00
funded by the Spanish
Ministry of Science and Innovation/State Agency of Research
MCIN/AEI/10.13039/501100011033 and by “ERDF A way of making Europe”.
\end{acknowledgements}

  \bibliographystyle{aa} 
  \bibliography{biblio} 

\begin{appendix}

\onecolumn
\section{Tables of ONC disks and JUMBOs}

\begin{table}[h]
    \caption{First 14 rows of coordinates, projected distance to ionizing source ($d_{\rm proj}$), Paschen $\alpha$ intensity and disk, IF radius and parallax from GAIA DR3 \citep{gaia2023_dr3} of the protoplanetary disks detected in the JWST NIRCam
 images of the ONC, combining GTO 1256 (GTO) and \textit{PDRs4All} (A,B Parallel, cf. Section~\ref{sect:data}) programs datasets. }
    \centering
    \begin{tabular}{ccccccccc}
    \hline
source & ra & dec &  $d_{\rm proj}$ & Pa $\alpha$ intensity & disk radius & IF radius & parallax\tnote{1} & dataset \\ 
        & (deg) & (deg) & (pc) &  $(\rm W~m^{-2}~sr^{-1})$  & (au) & (au) &  (mas)  &\\

\hline
\hline

005-514 & 83.7519933 & -5.4206395 & 0.50 & ... & ... & ... & 2.54±0.05 & GTO \\ 
006-439 & 83.7524771 & -5.4107850 & 0.47 & 8.32e-05±4.91e-06 & ... & ... & ... & GTO \\ 
016-149 & 83.7567717 & -5.3636212 & 0.46 & ... & ... & ... & ... & GTO \\ 
038-627 & 83.7674710 & -5.4410637 & 0.49 & ... & ... & ... & 2.52±0.06 & GTO \\ 
044-527 & 83.7684653 & -5.4243080 & 0.42 & 3.35e-05±2.18e-06 & ... & ... & ... & GTO \\ 
046-245 & 83.7693507 & -5.3791266 & 0.34 & 3.65e-05±2.85e-06 & ... & ... & ... & GTO \\ 
049-143 & 83.7706501 & -5.3619533 & 0.38 & 2.32e-05±1.61e-06 & ... & ... & ... & GTO \\ 
057-419 & 83.7738944 & -5.4051552 & 0.32 & ... & ... & ... & 2.63±0.06 & GTO \\ 
061-401 & 83.7754257 & -5.4001717 & 0.30 & 1.11e-05±2.64e-06 & ... & 85±12 & ... & GTO \\ 
066-652 & 83.7774933 & -5.4477779 & 0.49 & ... & ... & ... & ... & GTO \\ 
069-601 & 83.7788126 & -5.4335095 & 0.40 & ... & ... & ... & 2.66±0.15 & GTO \\ 
072-135 & 83.7800808 & -5.3595731 & 0.33 & ... & 113±4 & 110±12 & ... & GTO \\ 
073-227 & 83.7802916 & -5.3740444 & 0.28 & ... & ... & ... & 2.37±0.07 & GTO \\ 
090-326 & 83.7876288 & -5.3906216 & 0.21 & ... & 47±6 & ... & ... & GTO \\ 
\multicolumn{9}{c}{...} \\
\hline    
    \end{tabular}
    \tablefoot{The full table is available in electronic version at the CDS.}
 \label{tab:coordinates}
\end{table}

\begin{table}[h]
         \caption{Coordinates of JuMBO candidates reported by \citet{pearson_jupiter_2023}, contained in the \textit{PDRs4All} NIRCam images.  The coordinates are those derived in the NIRCam \textit{PDRs4All} images.}
    \centering

    \begin{tabular}{cccc}
    \hline
      Sources   &  \multicolumn{2}{c}{Position ra, dec (deg)}          & Module \\
                &  a                      &    b           &   \\ \hline\hline
    JuMBO24     & 83.8312915, -5.3943755  &                           & B \\
    JuMBO25     & 83.8364742, -5.3711567  & 83.8363771, -5.3710603    & A \\
    JuMBO26     & 83.8380085, -5.3665563  & 83.8378384, -5.3665275    & A \\
    JuMBO27     & 83.8466416, -5.3995442  & 83.8468808, -5.3995483    & B \\
    JuMBO28     & 83.8469629, -5.3927260  &                           & B \\
    JuMBO29     & 83.8472692, -5.3466808  & 83.8473650, -5.3466847    & A \\
    JuMBO30     & 83.8485400, -5.4059630  &                           & B \\
    JuMBO31     & 83.8567525, -5.3879075  & 83.8567387, -5.3877558    & B \\
    JuMBO36     & 83.8788150, -5.3402700  & 83.878595 , -5.3401447    & Parallel B \\
    JuMBO37     & 83.8822620, -5.3307497  & 83.8824791, -5.3307117    & Parallel B \\
    JuMBO38     & 83.8832808, -5.3519342  & 83.8831334,  -5.3519777   & Parallel B \\
\hline    
    \end{tabular}
    \tablefoot{
    The third column corresponds to the FOV in which the source is observed (module A or B, parallel, cf. Section~\ref{sect:data}).}

    \label{tab:coordinates_JuMBOs}
\end{table}

\onecolumn

\section{Maps of specific lines}\label{sect:All_maps}

\begin{figure*}[h]
  \centering
   \subfloat{
     \includegraphics[width=0.37\linewidth]{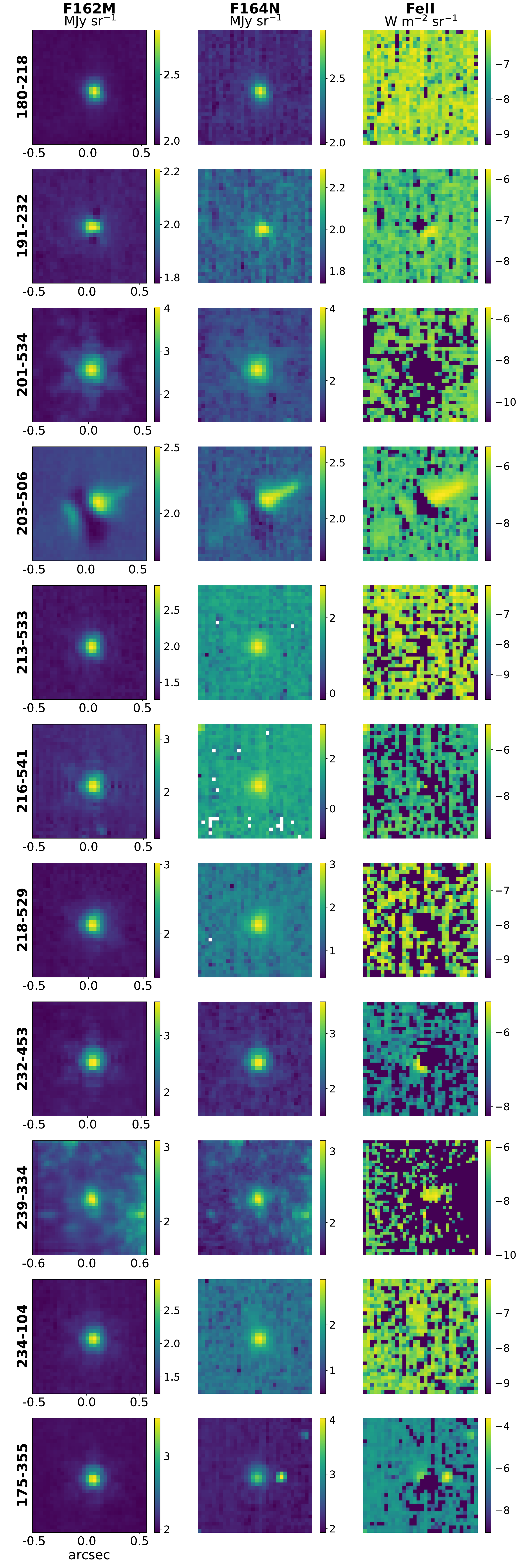}
   }
   \subfloat{
     \includegraphics[width=0.37\linewidth]{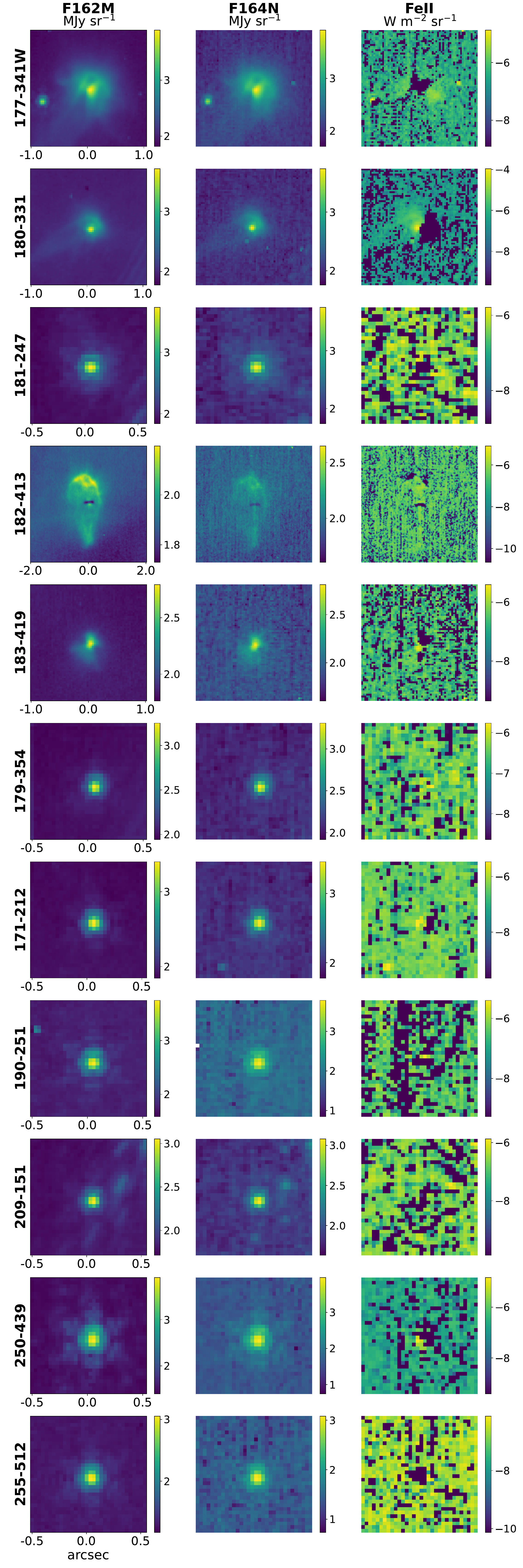}}
  \caption{[\ion{Fe}{ii}] emission map of 22 protoplanetary disks using \textit{PDRs4All} NIRCam images. The first and second column correspond to the image in the F162M and F164N filter, respectively. The third column is the [\ion{Fe}{ii}] emission map obtained following equation \ref{eq:recipes_lines}. The images are north-east aligned. 
  }
  \label{fig:FeII_images}
\end{figure*}

\begin{figure*}
  \centering
   \subfloat{
     \includegraphics[width=0.42\linewidth]{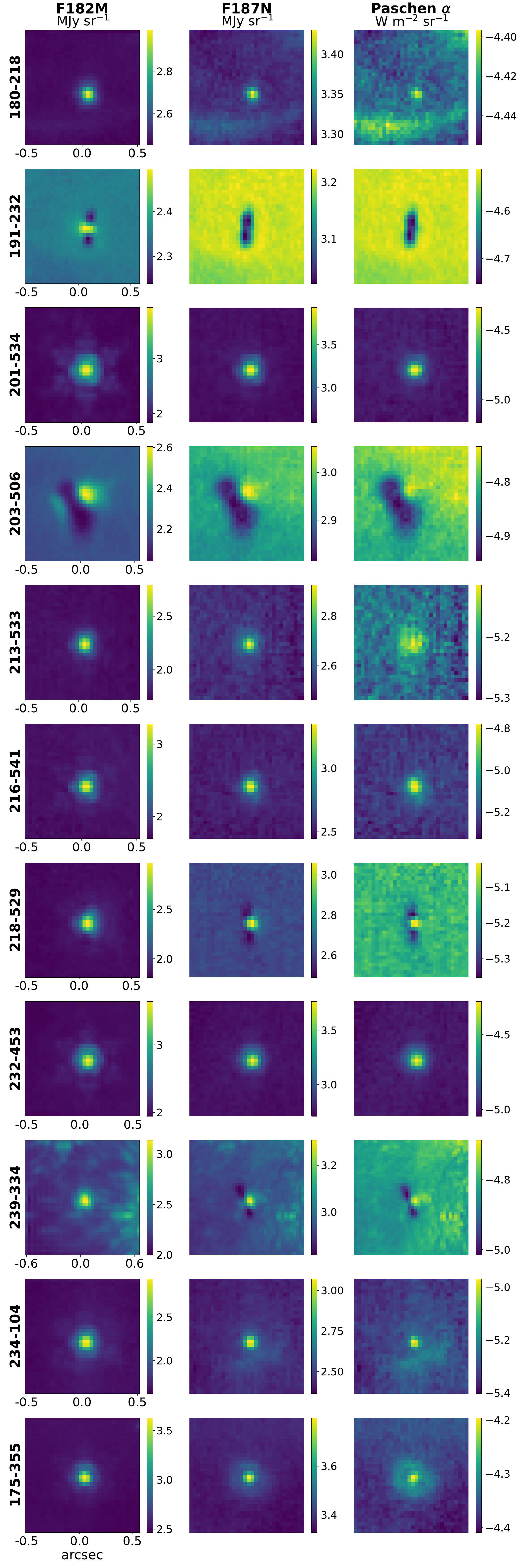}
   }
   \subfloat{
     \includegraphics[width=0.42\linewidth]{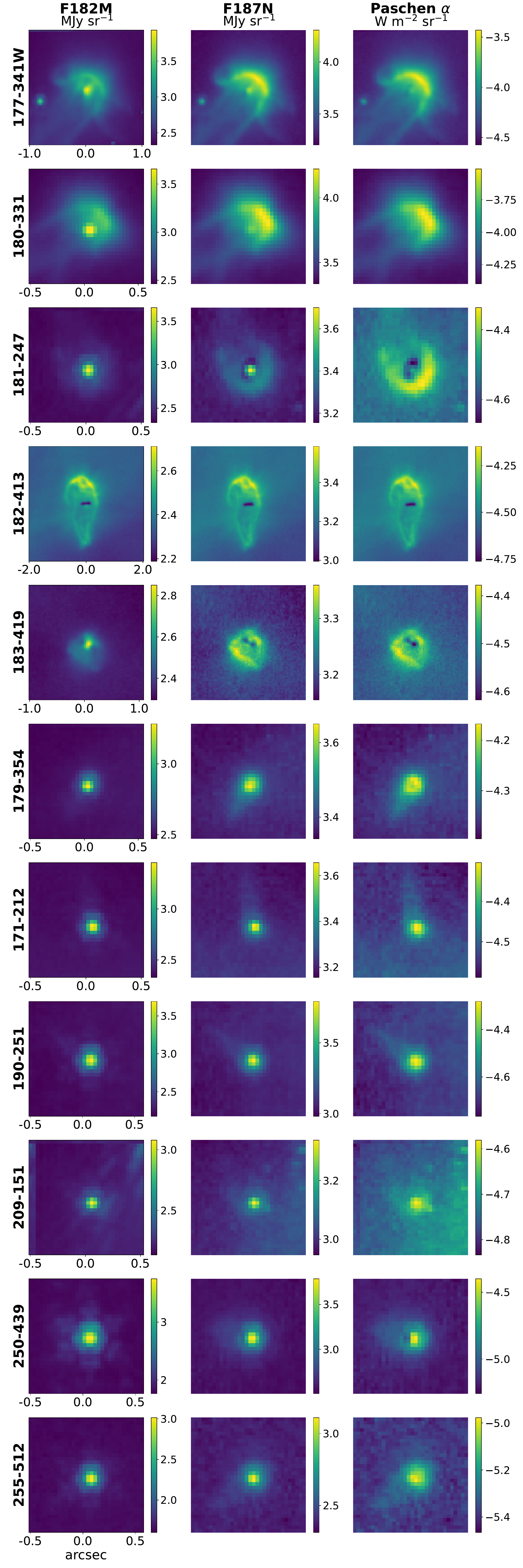}
  }
  \caption{Paschen $\alpha$ maps of 22 protoplanetary disks. 
  The first and second column correspond to the image in the F182M and F187N filter, respectively. The third column is the Paschen $\alpha$ emission map obtained following equation \ref{eq:recipes_lines}. The images are north-east aligned. 
  }
  \label{fig:Paschen_images}
\end{figure*}

\begin{figure*}
  \centering
   \subfloat{
     \includegraphics[width=0.42\linewidth]{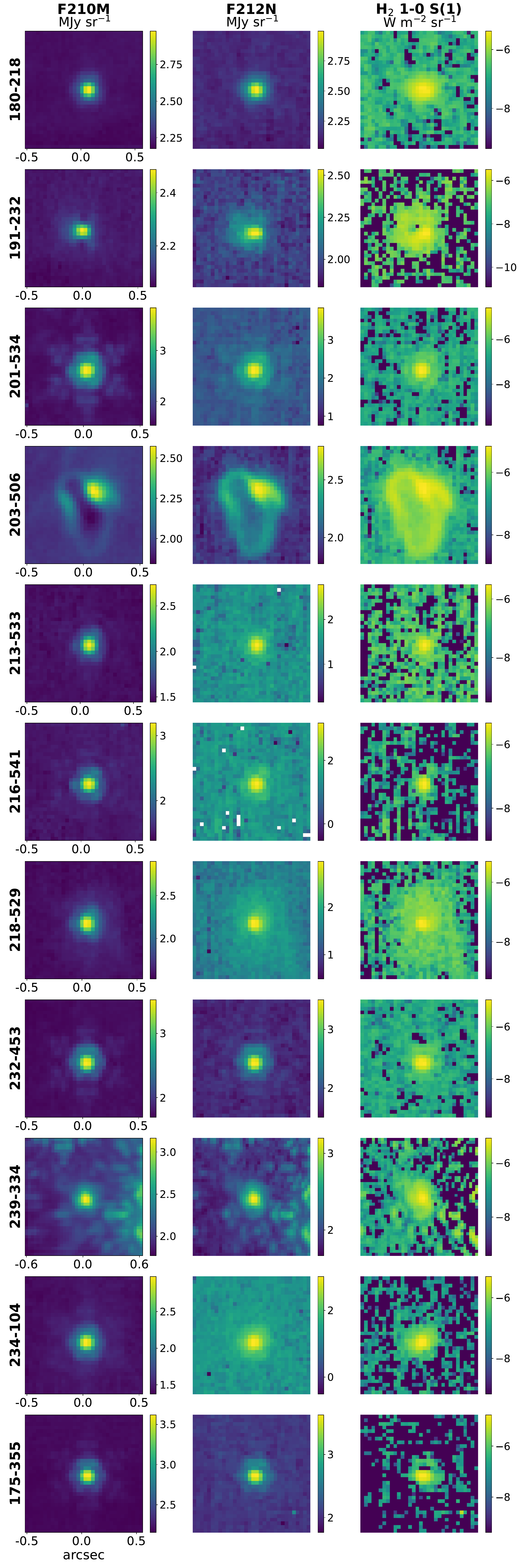}}
   \subfloat{
     \includegraphics[width=0.42\linewidth]{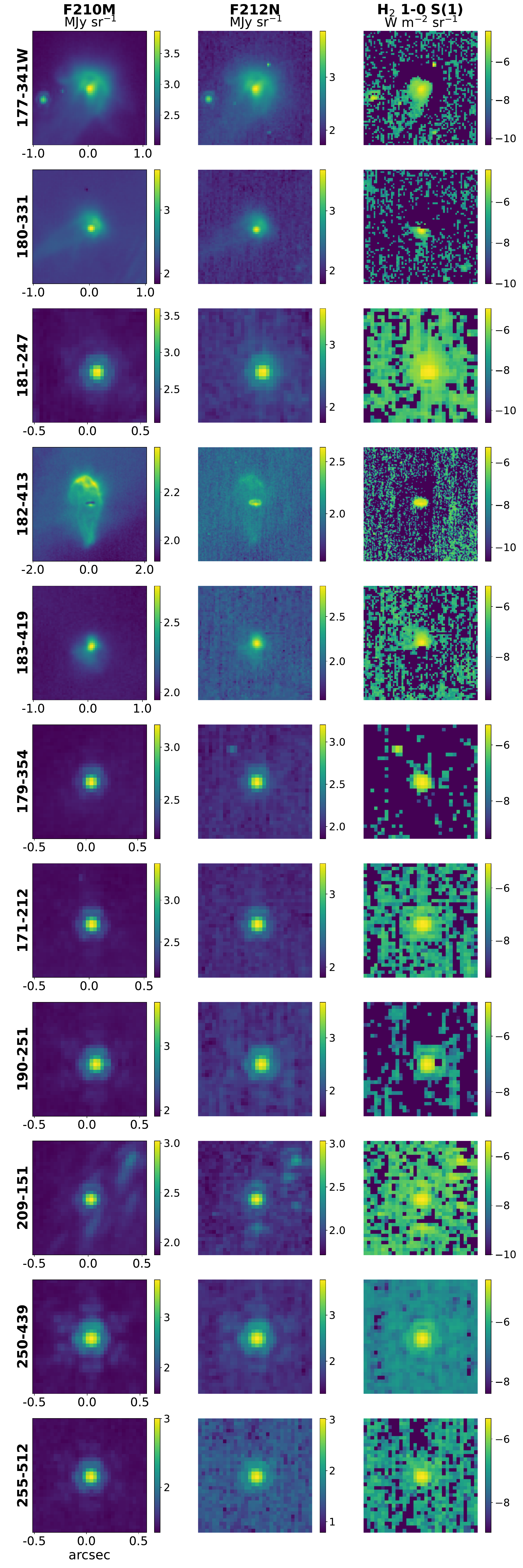}
  }
  \caption{H$_2$ 1-0 S(1) emission maps of 22 protoplanetary disks. 
  The first and second column correspond to the image in the F210M and F212N filter, respectively. The third column is the H$_2$ emission map obtained following equation \ref{eq:recipes_lines}. The images are north-east aligned.
  }
  \label{fig:H2_images}
\end{figure*}

\begin{figure*}
  \centering
   \subfloat{
     \includegraphics[width=0.42\linewidth]{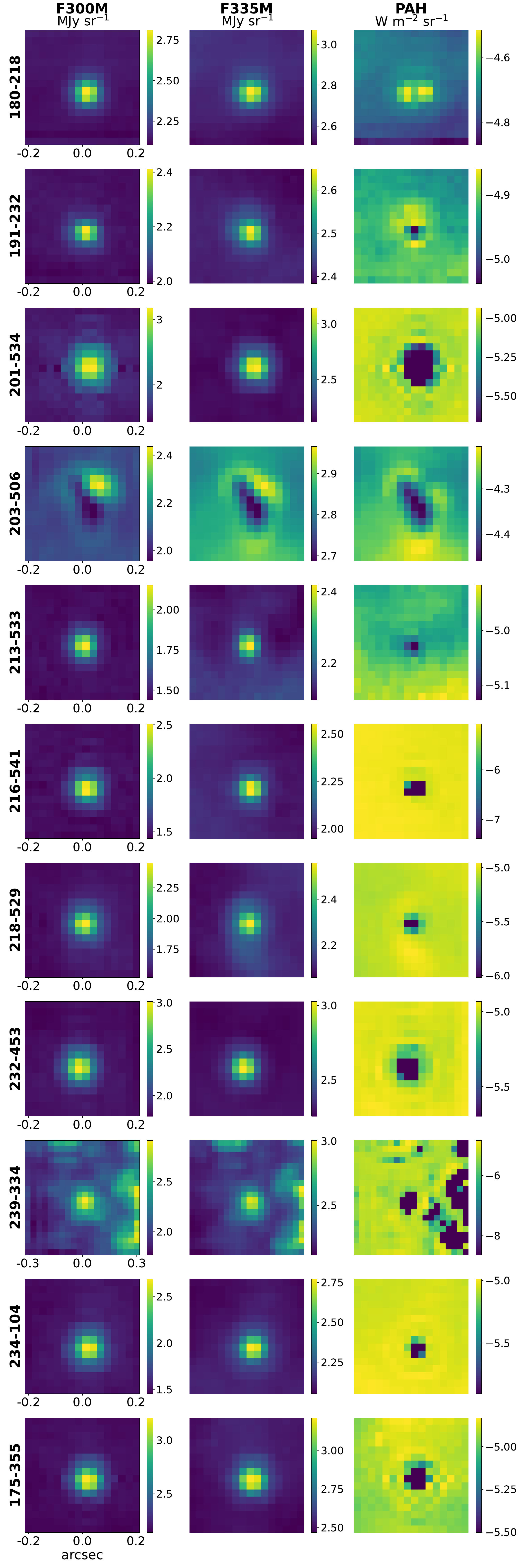}
   }
   \subfloat{
     \includegraphics[width=0.42\linewidth]{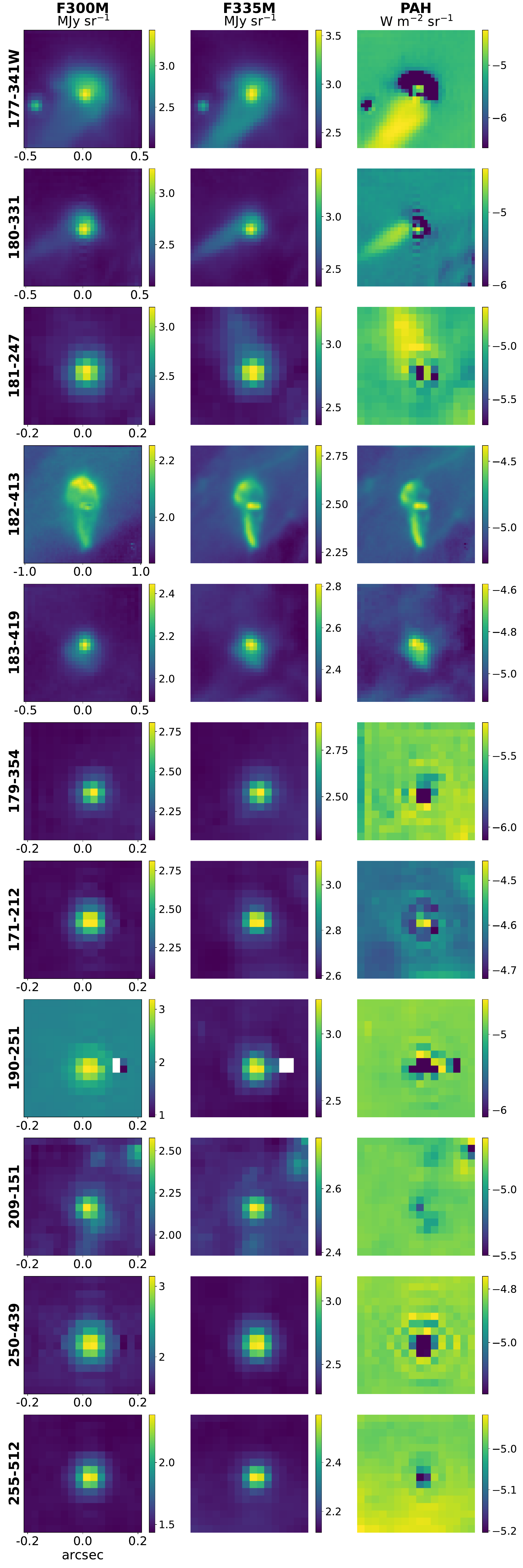}
  }
  \caption{PAH emission maps of 22 protoplanetary disks. 
  The first and second column correspond to the image in the F300M and F335M filter, respectively. The third column is the PAH emission map obtained following equation \ref{eq:recipes_lines}. The images are north-east aligned.
  }
  \label{fig:PAH_images}
\end{figure*}

\begin{figure}[h]
\section{NIRCam images of JuMBOs}
\centering    
     \includegraphics[width=0.91\linewidth]{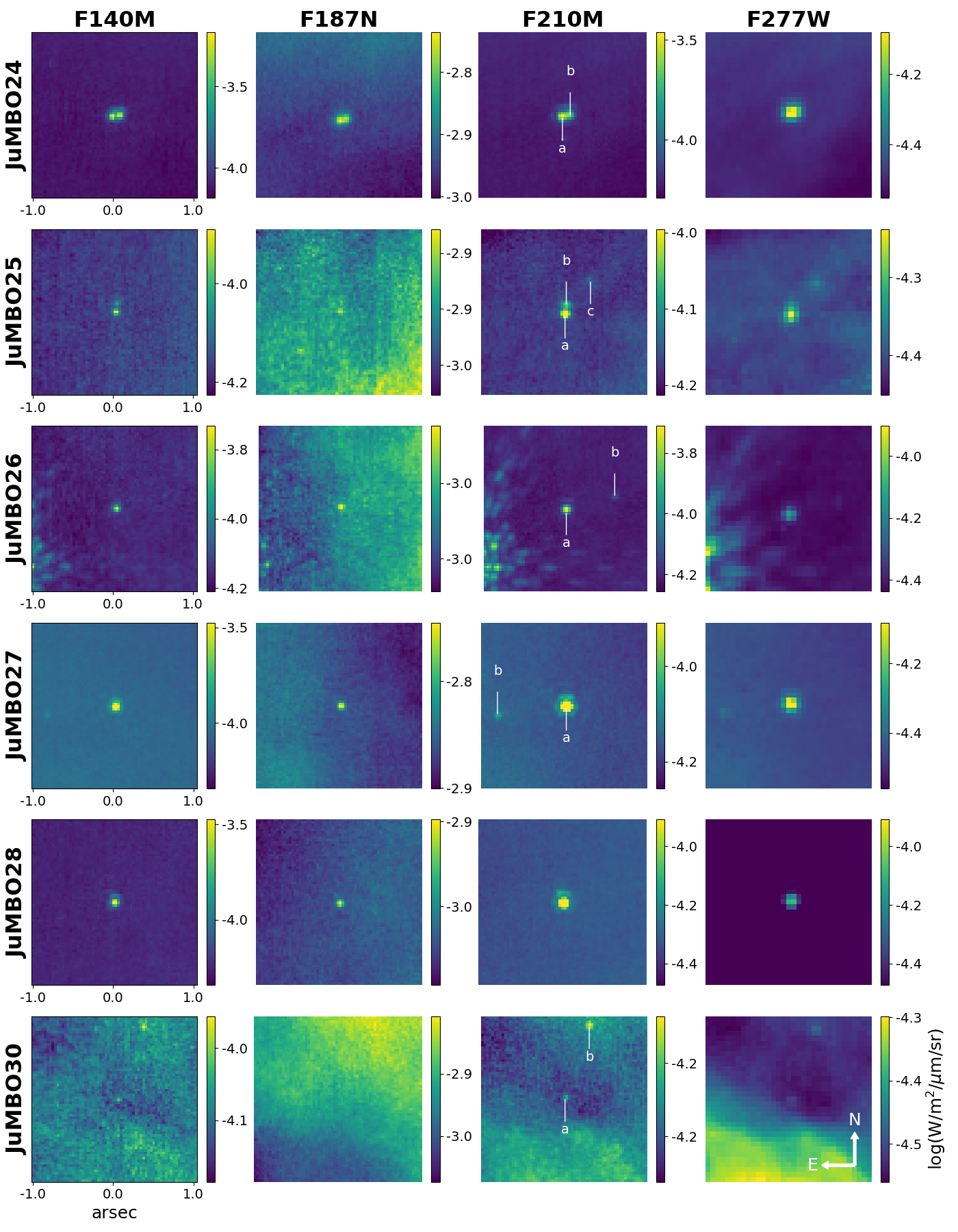}
     \caption{NIRCam images of six JuMBOs as seen in the \textit{PDRs4All} program dataset. The components of each systems are labeled a, b, c in the F210M images when at leat two sources are detected.
     The colorbars are in $\log (\rm W\:m^{-2}$ \textmu m$\rm^{-1}\:sr^{-1})$.}
     \label{fig:kaleidoscope_JuMBOs}
\end{figure}

\begin{figure*}
\section{Spectral energy distributions}
    \centering
    \includegraphics[width=\linewidth]{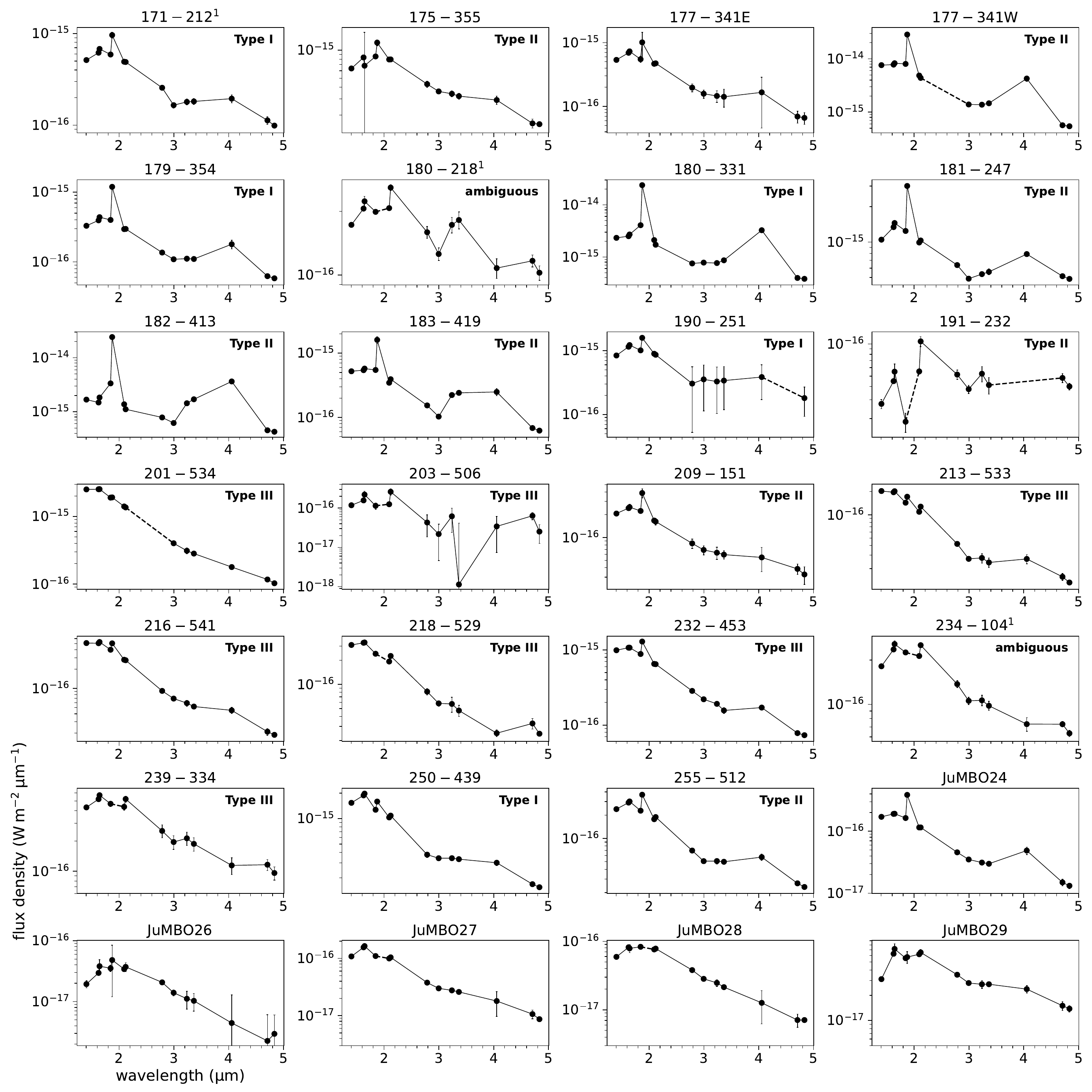}   
    \caption{Spectral energy distribution of the 23 protoplanetary disks and five JuMBOs in the \textit{PDRs4All} NIRCam images, obtained as described in Section~\ref{sect:SED}. 
    The dashed segments in some of the SEDs replace saturated filters. The $3\sigma$ error bars are given.
    The type of the source (type I, II or III, cf. Section~\ref{sect:typology}) is indicated in the upper right corner when available.\\ $^{1}$: Proplyd candidates from \citet{habart_2024_NIRCam}. }
    \label{fig:SED_proplyds}
\end{figure*}

\FloatBarrier

\twocolumn

\section{NIRCam image of HC182}
\begin{figure}[h]
    \centering
    \includegraphics[width=0.8\linewidth]{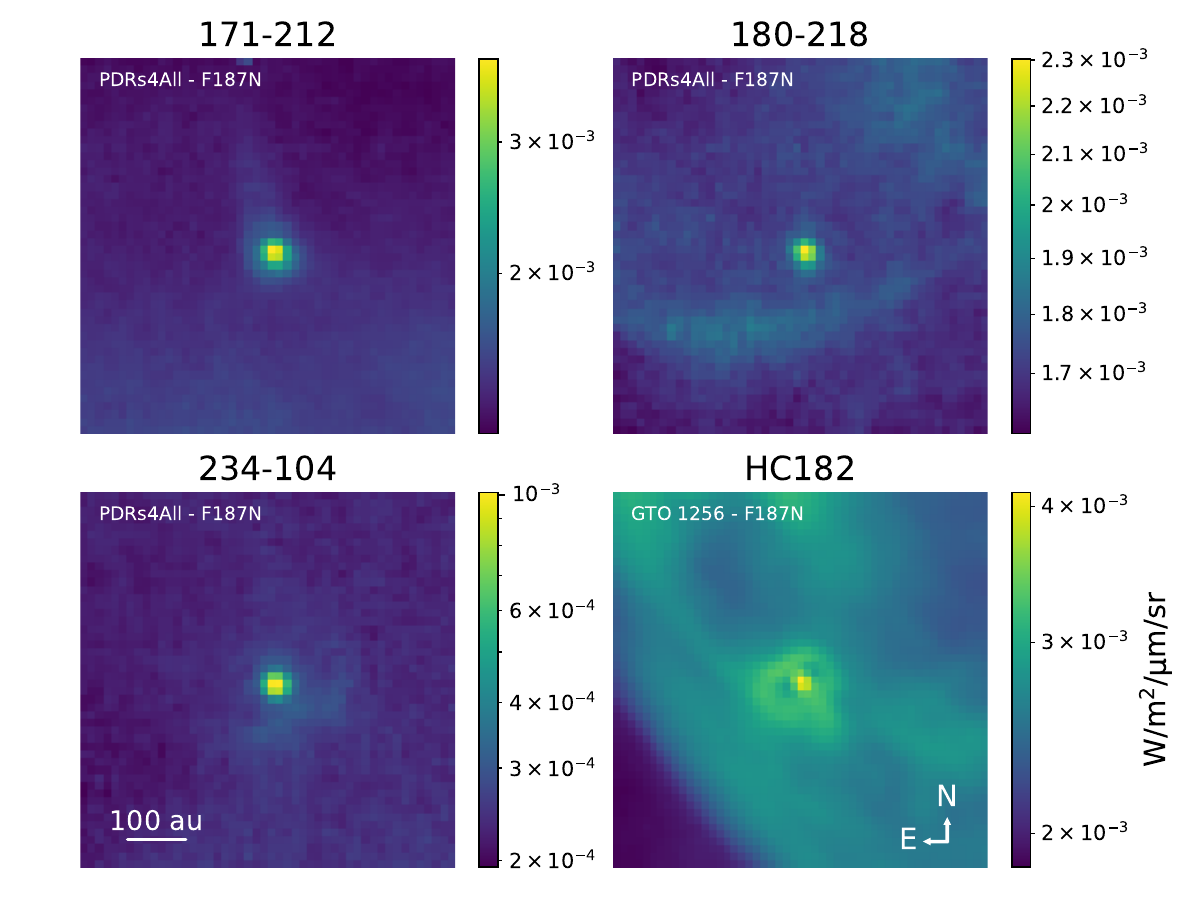}
    \caption{NIRCam image in the F187N filter of the three proplyd candidates reported by \citet{habart_2024_NIRCam} from the \textit{PDRs4All} program, and of HC182 from the JWST GTO 1256 program.}
    \label{fig:HC182}
\end{figure}

\section{JuMBO24}
\subsection{Individual short wavelength SEDs of the two components of JuMBO24}\label{sect:Appendix_JuMBO24}

\begin{figure}[h]
  \centering
     \includegraphics[width=0.6\linewidth]{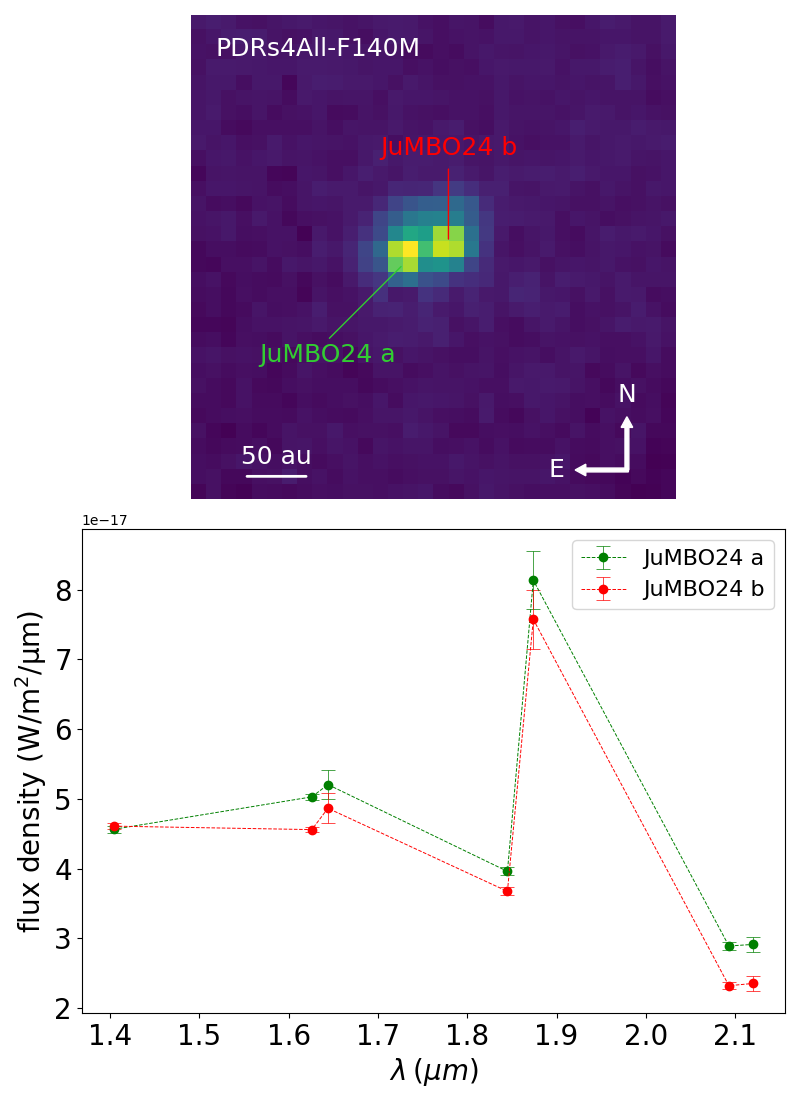} 
  \caption{Top panel: F140M images of JuMBO24 a and JuMBO24 b. Bottom panel: Individual short wavelength SEDs of JuMBO24 a (green) and b (red).}
  \label{fig:JuMBO24_SWL}
  \end{figure}

Fig.~\ref{fig:JuMBO24_SWL} shows the F140M image of JuMBO24 a and JuMBO24 b (top panel), and their associated SED in the short wavelengths channel of NIRCam (bottom panel), where the binary system is resolved .
The SEDs are obtained using the method presented in Section~\ref{sect:SED} with 1.5-pixel-radius apertures to minimize the contamination from the companion.

\subsection{JuMBO24 SED fit}

We use the \textsc{Phoenix} model (parameters: $T_{\text{eff}}$, $\log g$, metallicity) combined with a blackbody ($T_{\text{disk}}$) and two Gaussians with $\sigma =10^{-3}$~\textmu{}m at 1.87 and 4.05~\textmu{}m, respectively, to model the SED of JuMBO24. We assumed solar metallicity and $\log g=3.0$ ($g$ in $\rm cm~s^{-2}$).
The effect of dust extinction on the line of sight, parametrized by the visual extinction $A_{\rm V}$, is considered using the extinction curve of the model of \citet{Weingartner_Draine_2001} with $R_\text{V} = 5.5$, {the value established for the ONC {\citep[e.g.,][]{Cardelli_1989, Fang_2021}}}.
The model is computed over a grid spanning an effective temperature $T_{\text{eff}}$ from 2300~K to 3500~K with a step size of 100~K,  disk temperature from 200~K to 1200~K with a 100~K step size, and visual extinction $A_{\rm v}$ from 0 to 5.0 in 0.5 steps. 
We finally keep the set of parameters that minimizes the reduced  $\chi^2$ fit to the data.
The obtained best-fit model is shown in Fig.~\ref{fig:JuMBO24_SED_fit}.

\section{Emission measure from Paschen $\alpha$ and radio emission a}\label{sect:EMradio}

At the IF of an EUV irradiated disk, electron recombination producing the Pa$\alpha$ intensity scales with the emission measure $EM$ \citep{balog_photoevaporation_2008} following $I_{\rm Pa\alpha}=2.51\times10^{-19}EM$,
where $I_{Pa\alpha}$ is the Pa $\alpha$ intensity in $\rm erg\ s^{-1}\ cm^{-2}\ arsec^{-2}$ and $EM$ is in $\text{cm}^{-6}\text{pc}$. 
To obtain $I_{\rm Pa\alpha}$ we divide the Pa$\alpha$ flux density for one object by the surface area of the aperture photometry (cf. Section~\ref{sect:SED}).
This yields $EM_{\rm Pa\alpha} = 5.8 \times 10^{6}$ cm$^{-6}$ pc.

\citet{rodriguez_radio_2024} reported radio continuum emission at 6.1 and 10 GHz arising from both components of JuMBO24 with archival Karl G. Jansky Very Large Array (VLA) observations. Assuming this emission originates from free-free radio emission at the IF, we can derive the emission measure ($EM$) following \citet{mezger_henderson1967}:
\[
\dfrac{EM}{\rm cm^{-6}\ pc} = \left( \dfrac{\tau_{\rm ff}}{3.28\:10^{-7}}  \right) \left( \dfrac{T_e}{10^{4}\ \rm K}  \right)^{1.35} \left( \dfrac{\nu}{\rm GHz}  \right)^{2.1}\ ,
\]
with $\tau_{\rm ff} = \dfrac{I_\nu}{B_\nu(T_e)}$ \citep{ballering_isolating_2023} the optical depth for free-free emission and $T_e \sim 10^4$K the electron temperature. 
At a frequency $\nu = 10 \:$ GHz, \citet{rodriguez_radio_2024} derived a flux for JuMBO24 of $F_\nu=60\pm13 \ \mu \text{Jy}$. 
With a beam size of $0.21"\times0.17"$ at this frequency, this yields a specific intensity:
\[ I_\nu = \dfrac{F_\nu }{\rm Beam} = (7.2\pm 1.5)\:10^{-19} \: {\rm W\ m^{-2}\ Hz^{-1}\ sr^{-1}}. \] 
The optical depth is thus $\tau_{\rm ff} = (2.3\pm0.5)\times 10^{-3}$, giving an Emission Measure $EM_{\rm radio} = (9.0\pm2.0)\:10^{5}\ {\rm cm}^{-6}\ {\rm pc}$, where the uncertainty corresponds to the error on the flux.
The emission measure obtained with the VLA radio continuum emission and with the Pa $\alpha$ NIRCam intensity are of the same order of magnitude. This indicates that the observed radio emission can be explained by the presence of an IF.

\section{Photon luminosity of $\theta^1$ Ori C}\label{sect:thetaOriC}

The main ionizing source in the ONC is $\theta^1$ Ori C \citep{simondiaz_2006_trapezium}. We modeled its spectrum using the OSTAR2002 grid \citep{Lanz_2007} considering $\log g = 4.1$, $R = 10~\rm R_\odot$, $T=4\times10^4~\rm K$, $Z=1$, which are the parameters estimated by \cite{simondiaz_2006_trapezium}. 
The EUV photon luminosity $\Phi_{\rm EUV}$ corresponds to the integrated photon flux for wavelengths below the Lyman limit at 912~\AA. We obtain $\Phi_{\rm EUV}=1.29\times10^{49}~ \rm s^{-1}$, {close to to the value provided by \citet{tielens_2005book} ($1.3\times 10^{49}~\rm s^{-1}$) for an O7-star such as $\theta^1$ Ori C \citep{simondiaz_2006_trapezium}. This value is slightly below the value provided by \citet{aru_kaleidoscope_2024} ($\Phi_{\rm EUV}=1.6\times 10^{49}~\rm s^{-1}$), but larger than the one obtained by \citet{odell_2017_whichstar} ($\Phi_{\rm EUV} = 7.35\times 10^{48}~\rm s^{-1}$)}.

The FUV photon luminosity is the flux integrated in the range 912-2400~\AA. This provides $\Phi_{\rm FUV}=3.49\times 10^{49}~ \rm s^{-1}$.
The fraction of FUV photons with respect to EUV photons is then $f=\Phi_{\rm FUV} / \Phi_{\rm EUV}=2.7$, {close to the value provided by \citet{ferland_2012} of 2.76}.

\FloatBarrier 

\end{appendix}

\end{document}